\documentclass[review]{elsarticle}

\usepackage{hyperref}
\usepackage{subfigure}
\usepackage{amsmath}
\journal{Journal of Nuclear Physics B}
\begin{document}

\begin{frontmatter}

\title{{\bf Planck-scale induced left-right gauge theory at LHC  and  
experimental tests}}
\author{\bf M.K. Parida$^{*}$ and  Biswonath Sahoo$^{\dagger}$ \\
\fnref{myfootnote}}

\address{Centre of Excellence in Theoretical and Mathematical Sciences, \\
Siksha 'O' Anusandhan University, Khandagiri Square, Bhubaneswar 751030, \\
Odisha,  India\\}

\cortext[mycorrespondingauthor]{Corresponding author}
\ead{minaparida@soauniversity.ac.in}

\begin{abstract}
Recent measurements at LHC has inspired searches for TeV scale left-right gauge
theory originating from grand unified theories. We show that inclusion of
Planck-scale induced effects due to ${\rm dim.}5$ operator not only does away with all the additional intermediate
symmetries, but also it predicts the minimal set of light Higgs scalars
tailored after neutrino masses and dilepton, or trilepton signals. The
heavy-light neutrino mixings are predicted from charged fermion mass fits in
$SO(10)$ and LFV constraints which lead to new predictions for dilepton
or trilepton production signals. Including 
fine-structure constant matching and two-loop, and threshold effects predict
$M_{W_R}= g_{2R}10^{4.3\pm 1.5 \pm 0.2}$ GeV and proton lifetime
$\tau_p=10^{36.15\pm 5.8\pm 0.2}$ yrs with $W_R$ gauge boson coupling $g_{2R}=0.56-0.57$.
Predictions on lepton flavour and
lepton number violations are accessible to ongoing experiments.
Current CMS data on di-electron excess at $\sqrt s= 8$ TeV are found to be consistent with $W_R$ gauge boson mass $M_{W_R}\ge
1.9-2.2$ TeV which also agrees with the values obtained from dijet
resonance production data. We also discuss plausible explanations for
diboson production excesses observed at LHC and make predictions
expected at $\sqrt s =14$ TeV  
\end{abstract}

\begin{keyword}
\texttt{left-right symmetry, D-parity, Grand unification, LHC} 
\end{keyword}

\end{frontmatter}


\section{Introduction}

 The standard model
 $SU(2)_L\times U(1)_Y\times SU(3)_C ~(\equiv G_{213})$ 
partially unifies electromagnetic and weak interactions but fails to explain
neutrino masses and why parity violation occurs only in weak interaction. 
Manifestly left-right symmetric (LRS) gauge theory \cite{rnmrev,pati,rnm,rnmgs:1981}
$SU(2)_L\times SU(2)_R \times U(1)_{B-L} \times SU(3)_C 
(g_{2L}=g_{2R}) (\equiv G_{2213D})$
predicts a number of phenomena beyond the standard model including neutrino
masses and parity violation. It also goes further to suggest that the
right-handed (RH) neutrino ($N$), a member of its fundamental
representation, could be a heavy Majorana fermion driving type-I seesaw
mechanism for light neutrino masses and acting as a seed for
baryogenesis via leptogenesis. As possible experimental evidence of LRS theory, it
would be quite attractive
to associate these RH neutrinos to be mediating dilepton
production events recently observed at the Large Hadron Collider (LHC)
\cite{cms1,cms2} which can discriminate whether $W_R$ gauge coupling
is different from the standard $W_L$ boson coupling \cite{cmp}.

There are a number of advantages of embedding the SM or the LRS models in
GUTs which have attracted extensive invesigations over the last four
decades \cite{pati,cmp,Georgi:1974,RS:1981,RS:1982}. The most recent phenomena have been
the prediction of dark matter (DM) candidates
including the stabilising symmetry,  called the
Matter Parity, in non-SUSY SO(10) \cite{khp:2011}. In addition to unifying the strong, weak,
and electromagnetic forces, the grand unified theory (GUT) is capable
of addressing the issue of
proton stability, and the origin of Parity and CP symmetries as part
of gauge symmetries.

The minimal left-right symmetric GUT that unifies strong, weak, and
electromagnetic interactions is $SO(10)$ that leaves out gravity \footnote{In
  the absence of any experimental evidence of supersymmetry so far, in
  this work we
  confine to non-supersymmetric (non-SUSY) models.}. Apart from fitting all charged
 fermion masses\cite{baburnm} and explaining the neutrino oscillation data,
it would be quite interesting if spontaneous symmetry breaking of
non-SUSY $SO(10)$ through any one of the following two minimal
symmetry breaking chains gives the LHC verifiable $W_R,Z_R$ bosons as
well as the associated seesaw mechanism

\begin{eqnarray}
{\bf SO(10) 
\displaystyle_{\longrightarrow}^{M_U}\,\,G_{2213D}~~~~ or~~~~G_{2213} 
\displaystyle_{\longrightarrow}^{M_R}\,\,SM}. \label{minchain}
\end{eqnarray}
  In eq.(\ref{minchain}) $G_{2213}$ represents the same left-right gauge theory
 as in $G_{2213D}$ but without the D-parity for which $g_{2L} \neq
 g_{2R}$ \cite{cmp}. 

That the resonant $W_R$ production accompanied by heavy RH Majorana neutrino exchange would manifest in
like-sign dilepton signals at accelerator energies was suggested 
earlier \cite{ks}.
An interesting interpretation of the LHC data \cite{cms1,cms2} on the excess of events
in the like-sign dilepton channel $pp\to eejj$ along with the
reported ratio of $14:1$ of opposite sign to the same sign dilepton
signals has been made very recently in the context of minimal
left-right symmetric model (MLRSM) with $g_{2L}=g_{2R}$
\cite{gluza:2015} which has the  Higgs scalar bidoublet $\Phi(2,
2, 0, 1)$ and the triplets $\Delta_L(3,1,-2,1)\oplus \Delta_R(1,3,-2,1)$ 
\cite{rnmgs:1981}. The light neutrino mass matrix in this theory 
\cite{gluza:2015} is governed by
the type-I seesaw formula 
\begin{eqnarray} 
{\cal M}_{\nu}=-M_D{\tilde M}^{-1}_NM_D^T. \label{type-I}
\end{eqnarray}
Here $M_D=$ Dirac neutrino mass matrix, ${\tilde M}_N=fV_R=$  the
RH neutrino mass matrix, $f=$ Majorana type Yukawa coupling of the triplets
, and $V_R=<\Delta_R^0>$ that breaks MLRSM to SM.
 There are several limitations of deriving this TeV scale MLRSM 
 from SO(10):
(i) It was noted \cite{RS:1981,RS:1982} that when
the GUT symmetry breaking proceeds through MLRSM, low-mass parity restoration with $M_{W_R}\sim {\cal
  O}(100-1000)$ GeV needs too large
value of  $\sin^{2}\theta_{W}(M_Z)\sim~0.27-0.31$ in direct conflict with the current electroweak
precision data. It was also observed that the value of
$\sin^{2}\theta_{W}(M_Z)\sim~0.23$ pushes the $W_R$ mass prediction
in this minimal scenario to very large value $M_R> 10^9$
GeV.
In fact  the globally accepted values of
$\sin^{2}\theta_{W}(M_Z)\sim~0.23126\pm 0.00005$ and
$\alpha_S(M_Z)=0.1187\pm 0.0017$ \cite{pdg,PDG:2014} restrict the MLRSM 
intermediate breaking scale to
be large $M_R \ge 10^{10}$ GeV  \cite{lmpr} but with experimentally
acceptable proton lifetime.  
Thus the SO(10) origin of TeV scale MLRSM is ruled out by RG
constraints on gauge coupling unification.
(ii) The second limitation is imposed by the neutrino oscillation data
and their type-I seesaw embedding in SO(10). The underlying quark-lepton symmetry \cite{pati} in SO(10)
predicts $M_D \sim M_u$ where $M_u=$ up-quark mass matrix. Then the
 explanation of neutrino oscillation data through 
 eq.(\ref{type-I}) predicts the seesaw scale to be too large, $M_R=10^{11}\to 10^{14}$ GeV ruling out any prospect of direct verification of SO(10) based MLRSM or type-I seesaw at accelerator energies.
(iii) Even if the TeV scale $G_{2213D}$ symmetry is shown to emerge from SO(10) by severely relaxing the ESH as in ref. \cite{lindner:1996}
discussed below, it may also have the cosmological domain wall problem\cite{Kibble:1980,Kibble:1982,Dborah:2011,Kuchimanchi:2012}. 
The resulting
massive domain wall would  contribute to mass density of
the universe upsetting the observed values. This calls for
adopting inflationary model of the universe which, however, is capable of
removing such a domain wall if the parity breaking scale is far above the
TeV scale. On the other hand with TeV scale paritry breaking,
the imposition of inflation and reheating at lower scale may not effectively remove
the domain wall. 
 
In the non-minimal LRS model with
$g_{2L}=g_{2R}$ consistent with
 the electroweak precision data, low scale $W_R,Z_R$ bosons have been
 realised, but this needs  unusually
larger number of nonstandard Higgs scalars and/or exotic fermions
\cite{lindner:1996} which drastically violate 
the ESH\cite{del}. Also no ansatz for neutrino oscillation data or LHC data
  have been
provided in this model.
This nonminimal model may also have the domain wall
problem as in the case of MLRSM discussed above.

On the other hand, the $G_{2213}$ model with high D-Parity breaking
scale  resulting in  $g_{2L} \neq g_{2R}$ at lower scales is free from
the domain wall problem \cite{cmp}.   
But  even  when the GUT symmetry breaks through the minimal $G_{2213}$, the
allowed solutions for TeV scale $W_R$ have been shown to require also a number of additional light particle
degrees of freedom \cite{malinsky:LR13}, although less than the nonminimal
$G_{2213D}$ case \cite{lindner:1996}. In this case also the ESH has to be 
abandoned. Further as in the case of ref.\cite{lindner:1996}, the glaring issue of neutrino
masses and mixings in these models  
  \cite{malinsky:LR13}  has not been addressed in direct contravention of the neutrino oscillation data
, let alone the LHC anomalies.\\

Although several possibilities have been discussed
earlier\cite{cmp,cmpgm,berto:2006}, in addition to preserving the
interesting property of fitting charged fermion masses,  
allowed solutions for TeV scale $W_R,Z_R$ bosons in the best identified chain
of ref.\cite{cmpgm} have been noted recently to be in concordance with the
neutrino oscillation data \cite{app}. This model has non-minimal
number of four intermediate symmetries instead of single LR
intermediate gauge theory at the LHC scale
\begin{eqnarray}
{\bf SO(10) \displaystyle_{\longrightarrow}^{M_U}\,\,G_{224D}
\displaystyle_{\longrightarrow}^{M_P}\,\,G_{224}
\displaystyle_{\longrightarrow}^{M_C}\,\,G_{2213} 
\displaystyle_{\longrightarrow}^{M_R^+}\,\, G_{2113}
\displaystyle_{\longrightarrow}^{M_R^0}\,\,SM}. \label{intsyms}
\end{eqnarray}
In ~eq.(\ref{intsyms}) $G_{224D}$
denotes the Pati-Salam symmetry  
 $SU(2)_L\times SU(2)_R\times
SU(4)_C ~(g_{2L}= g_{2R})~(\equiv G_{224D})$ with left-right discrete
symmetry and  $G_{224}$ denotes the same gauge symmetry without the
D-Parity.\\ 

This model comprising of two-step breaking of $G_{2213}$ to SM was
originally proposed in ref.\cite{cmpgm} where the SO(10) Higgs representations 
${54}_H,{210}_H,{126}_H$ and ${10}_H$ were used to achieve the desired gauge hierarchy with low mass $W_R,Z_R$ bosons. With the further addition of ${16}_H$, a second ${10}_H$, 
and three additional fermion singlets, in addition to retaining the low mass 
$W_R,Z_R$ boson prediction,
this model was used to fit all charged fermion masses and obtain the $9\times 9$
neutral fermion mass matrix of eq.(\ref{matnn}) of Sec.2 given below 
while fitting the
neutrino oscillation data via TeV scale gauged inverse seesaw
formula  in ref.\cite{app}. 
While predicting LFV decays with branching ratios only few to four
orders smaller than the experimental limits, the model also predicted new dominant
contribution to double beta decays in the $W_L-W_L$ channel due to
sterile neutrino exchange closer 
to their experimental values.
  This model of ref.\cite{cmpgm,app,pas} has been recently
used to interpret the observed dilepton excess at CERN LHC in $pp\to eejj$ to be due to
$W_R$ mediation with
$M_{W_R}\sim 2$ TeV \cite{Deppisch:2014}. However the validity and further
confirmation of  the model requires detection of the $Z_R$
boson mass at $M_{Z_R} \le 2$ TeV at collider
energies. On the other hand $SO(10)$ embedding of single intermediate breaking of $G_{2213}$
 at TeV scale to SM predict $M_{Z_R}\ge 1.7 M_{W_R} \ge
3.4$ TeV if $M_{W_R}\ge 2.0$ TeV.

In view of the LHC capability to discriminate among
different models \cite{ks,agu-saav,cdm,lhc}, alternative theoretical explorations for GUT origins
of  LR models with parity restoration at low scales ($g_{2L}=g_{2R}$)
or at high scales resulting in TeV scale values $g_{2L}
   \neq g_{2R}$  having additional experimentally verifiable signatures
   would be interesting. 

Very recently, in an interesting development in single step
   breaking scenario,  TeV scale LR gauge theory has been derived 
   including the additional light Higgs scalar $\phi_S(1,3,0,8)\subset {210}_H$
   and non-standard fermion
   pairs $\Sigma_L(3,1,0,1)\oplus \Sigma_R(1,3,0,1)\subset {45}_F $
   under $G_{2213}$ \cite{psb-rnm:2015}.   
   The model has been shown to be consistent with
   neutrino oscillation data and observed excesses at LHC detectors on
   $W_R\to eejj$
   , $W_R\to jj$, $W_R\to WZ$, and $W_R\to WH$ production
   channels with $g_R=0.51$.
It has also  potential to explain dark matter and baryon asymmetry of
   the universe through leptogenesis, and the LHC cross section ratio
   for production of opposite-sign dileptons to like-sign dileptons.
   However, the model predicts large unification scale leading to 
 proton lifetime beyond the Super K. and Hyper K. \cite{sk-hk}
 limits. The presence of additional scalars and fermions can be also 
 tested at colliders including LHC.  

 Without using any GUT, but under the general assumption of the presence of
   TeV scale LR theory with
   $g_{2L}\neq g_{2R}$, it has been also shown how the current LHC data
   are explained with $M_{W_R}\simeq  1.8-2.0$ TeV and with
   $g_{2R}=0.5$ \cite{bogdan:2015}.\\

The SUSY grand desert models predict the GUT scale to be 
$M_U=2\times 10^{16}$ GeV by using the electroweak (EW) precision
values of electromagnetic fine structure constant
$\alpha(M_Z)={(127.9\pm 0.1)}^{-1}$ and either $\sin^2\theta_W(M_Z)$
or $\alpha_S(M_Z)$ \cite{langacker:1993}. Since the GUT scale is only
about two orders less than the Planck scale, effects of quantum
gravitational corrections treated to be 
induced by ${\rm dim}.5$ operator scaled by Planck mass has been
investiagated by a number of authors
\cite{langacker:1993,hill,shafi,cal,wil,hall:1993,ppm}. 
Particularly,
gravitational smearing effect on the precision value of
$\alpha_S(M_Z)$ was noted in ref.\cite{hall:1993} while it was shown
in ref. \cite{langacker:1993} that, in SUSY grand desert scenario, the
predicted value of any one of the two,  $\sin^2\theta_W(M_Z)$
or $\alpha_S(M_Z)$, is smeared out if the other is fixed at its EW precision value.  
Noting that such smearing effects due to quantum gravity is absent in
any intermediate scale model,   
the purpose of this work is to show that when Planck-scale induced effect is
included through a ${\rm dim}.5$ operator of the type dicussed earlier
\cite{langacker:1993, hill,shafi,cal,wil,hall:1993,ppm,skm:2007},
the $SO(10)$ model gives LHC scale LR gauge theory $G_{2213}$ 
in the minimal symmetry breaking chain with reduced  size
of the light Higgs spectrum consistent with gauged inverse seesaw
formula for neutrino masses that depends upon whether the RH neutrino masses 
are Pseudo-Dirac (Model-I) or Majorana (Model-II) fermions leading to the  manifestation of
$W_R$ through trilepton or dilepton signals at the LHC.
For the first time in the context of higher dimensional operator
effects, in addition to the analytic derivation of RGE's for $\ln
(M_U/M_Z)$, and $\ln (M_R/M_Z)$ , the third
RG equation 
is derived that ensures determination of the GUT coupling through
 electromagnetic fine-structure constant matching.
The model predicts
heavy-light neutrino mixings falling between two bench mark scenarios
\cite{agu-saav,cdm}
defined by the upper limit ${\vert V_{lN} \vert}^2=3\times 10^{-3}$ and the  
lower vanilla seesaw limit (${\vert V_{lN} \vert}^2=
{\sqrt (\Delta m^2_{\rm atm})}/M_N$) 
which constitute important ingredients for
dilepton or tri-lepton production signals at LHC detectors especially in the $W_L-W_L$ and $W_R-W_L$
channels and for the light sterile neutrino mediated
 $0\nu\beta\beta$ decay, and charged lepton flavor violating (LFV) branching ratios
closer to their experimental limits.
In the $RR$ channel, the Model-II explains the di-electron excess recently observed at the CMS detector \cite{cms2}
for $M_{W_R} = 1.9 -2.2$ TeV and both the models are found to explain
the dijet resonance data\cite{CMSdij,ATLASdij}, and excess of events
observed in the diboson production channels $W_R\to WZ$ and $W_R\to
WH$.
 We also make predictions for LHC run-II at $\sqrt s=14$ TeV in the
 $LL$, $RR$, and $RL$ channels for like-sign dilepton production cross
 sections.
 
   This paper is organized as follows. In Sec.2  we  discuss the
   predictions of $W_R$   and  the grand unification
   scales using the 
 ${\rm dim}.5$ operator. In Sec.3 we give a short description
on neutrino masses and LFV decay and in Sec.4 we discuss lepton number
violation. In Sec.5 we discuss how LHC provides hints for $W_R-$ boson
production in $pp$ collisions manifesting in dilepton and trilepton
signal cross sections. In Sec.6 we show how $W_R$ boson
mass is determined from the dijet resonance data while explaining
the diboson production data.
 Finally we give a brief summary of our results.

\section{LHC scale LR theory}

\subsection{\bf Planck-scale induced corrections to RG equations}
We attempt to predict the scale of LR gauge theory $G_{2213}$ 
in the minimal symmetry breaking chain of eq.(\ref{minchain}) while taking into account the Planck-scale induced corrections to RG equations for gauge couplings.
 We use the standard two-loop RG equations for gauge couplings 
 \begin{equation}
  \mu\frac{\partial g_{i}}{\partial \mu}=\frac{a_{i}}{16\pi^{2}}g_{i}^{3}+\frac{1}{(16\pi^{2})^{2}}\sum_{j}b_{ij}g_{i}^{3}g_{j}^{2}.
 \end{equation}
 We also include the effect of dim.5 operator which was first
 suggested in the context of SUSY $SU(5)$ \cite{hill} and non-SUSY
 $SO(10)$ with Pati-Salam intermediate symmetry \cite{shafi}, and
 subsequently used to examine modifications of various GUT predictions
 \cite{langacker:1993,cal,hall:1993,ppm,skm:2007,rizzo,pp}. In the absence of any specific well defined terms due to gravitational interaction, the ${\rm dim}.5$ operator scaled by the Planck mass has been treated to represent the effect of quantum gravity especially in SUSY SU(5) and its influence has been shown to smear out
the strong interaction coupling $\alpha_S(M_Z)$
\cite{langacker:1993,hall:1993}. Also such effects on GUT predictions
attributed due to quantum gravity effects have been investigated
further\cite{cal,wil}. In ref.\cite{shafi}, however, the effect has also been
attributed to be arising out of Kaluza-Klein type spontaneous
compactification of extra dimensions where the scale of the ${\rm
  dim.} 5$ operator could be lower than $M_{Planck}$.
 In our opinion the operator which is most
 effective in bringing
 $G_{2213}$  to $\sim {\cal O}$ (TeV) scale in single-step breaking of
 $SO(10)$ is 
 \begin{equation}
 \mathcal{L}_{NR}=\frac{C}{M_{Planck}} Tr(F_{\mu\nu}\phi_{(210)}F^{\mu\nu}),\label{kin}
\end{equation} 
 where $\phi_{210} \equiv 210_{H}$ Higgs representation that breaks $SO(10)\rightarrow G_{2213}$  at the GUT scale by acquiring 
 vacuum expectation value (VEV) along its $G_{2213}$ 
 singlet direction as defined below in eq. (\ref{vev210}) and  the scale of the operator is fixed at $M_{Planck}\simeq 2.4\times 10^{18}$ GeV, the reduced Planck mass. Because of the presence of intermediate symmetry, the gravitational smearing effects on $\alpha_S(M_Z)$ or $\sin^2\theta_W(M_Z)$, otherwise present in SUSY
grand desert models, are drastically reduced. 

It has been shown \cite{cmp} that there are two singlets in ${210}_H$ under LR
gauge group: the D-parity even singlet $\eta_e$ contained in Pati-Salam
sub-multiplet ${(1,1,15)}_{H}$ and the D-Parity odd singlet $\eta_o$
contained in Pati-Salam singlet ${(1,1,1)}_{H}$
of ${210}_H$.  
It was at first claimed \cite{rizzo} that when $SO(10)$ is broken
along the direction $<\eta_e> \sim M_{GUT}$, low-mass $W_R$ would
result through eq.(\ref{kin}) and one-loop contributions of certain light Higgs scalars . But it was noted \cite{pp}
that this solution is ruled out as it requires too large values
of $\sin^2\theta_W(M_Z)= 0.27$. Although accurate values of $\alpha_S(M_Z)$ and
$\sin^2\theta_W(M_Z)$ and neutrino oscillation data 
 were not available at that time, it was noted that \cite{pp} a low-scale $W_R$ would
require parity breaking at the GUT scale. Attempts have been made to
predict TeV scale LR gauge symmetry by using more than one intermediate
symmetry and through still higher dimensional operators which
introduce a number of additional parameters into the theory. We do
not discuss  them here as our aim here is to obtain the
LHC scale LR theory by direct breaking of $SO(10) \to G_{2213}$ in the
minimal chain with minimal number of parameters.

We note that eq.(\ref{kin}) is the only possible ${\rm dim}.5$ operator
that gives LHC scale $G_{2213}$ symmetry with minimum number of parameters,  
when the Higgs field $\Phi_{210}$ acquiring VEV along a 
direction which is a linear
combination of $<\eta_e>$ and $<\eta_o>$ defined through
eq.(\ref{vev210}) below. To understand this, we note that when  $\Phi_{210} \sim \eta_o$ in
eq.(\ref{kin}), we get only Pati-Salam symmetry and not
$G_{2213}$. Similarly when  $\Phi_{210} \sim \eta_e$  we get
$G_{2213D}$ with unbroken parity at a high scale ($M_R > 10^9$ GeV) with
$g_{2L}=g_{2R}$. Also when $\Phi_{210}$ is replaced by $\Phi_{45}$ that
contains the other D-odd singlet $\eta_o^{\prime}$,  eq.(\ref{kin})
vanishes identically. The only other possibility, besides the one used
here is to use  two different
${\rm dim}.5$ operators of the type  eq.(\ref{kin})  with two different coefficients where
in one operator $\Phi_{210}$ is aligned along $\eta_e$ and in the
other, it
 is aligned along $\eta_O$. This would introduce one additional
parameter compared to the present minimal model.

It is well known that in the absence of any threshold or higher dimensional operator effects, the two mass scales $M_U$ and $M_R$ in the single intermediate 
scale model are determined in terms of  $\alpha_S(M_Z)$ and $\sin^2\theta_W(M_Z)$ with
the fixed value of the fine-structure constant $\alpha_{(M_Z)}={1/127.9}$. 
 We show here analytically, 
 through three different new equations, how the
additional parameter due to eq.(\ref{kin}) changes the two mass scales
provided the GUT coupling is fixed by matching the 
electromagnetic 
fine-structure constant 
by the third equation which is an essential constraint in the model
in order to prevent any mismatch or gravitational smearing of $\alpha_(M_Z)={1/127.9}$ that may result due to such additional new corrections at the GUT scale. 
The fourth equation determines $C$ of eq.(\ref{kin}) in terms of the
model parameters unambiguously. 

Whereas in the earlier LR models  derived in one-step breaking of $SO(10)$
\cite{lindner:1996, malinsky:LR13}, the important questions of neutrino
masses, LFV, LNV, and LHC signatures of $W_R$ or $N$ have been left
out, in this work we have addressed these issues. 
The minimal sets of Higgs representation $(210_H\oplus16_H\oplus10_H)$(Model-I) or $(210_H\oplus126_H\oplus16_H\oplus10_H)$(Model-II) with the added presence
of three fermion singlets, not only makes the $W_R$, $Z_R$ bosons
accessible near the TeV scale, but also both the models are in concordance with the neutrino 
oscillation data through gauged inverse seesaw formula for neutrino
masses \cite{rnmvalle}  while predicting sizable charged lepton flavor
violating decays accessible to ongoing search experiments. This may be contrasted with all
previous dim.5 operator models in non-SUSY $SO(10)$ predicting
negligible lepton flavor violations. Exploiting the potential of
$SO(10)$ to yield Dirac neutrino mass matrix, our model predicts
heavy-light neutrino mixings used as important ingredients for
multi-lepton production signals at the LHC. 
While Model-I predicts trilepton
production decay signals mediated by the TeV scale Pseudo-Dirac neutrinos, Model-II predicts dominant dilepton signal accessible to LHC mediated by RH
Majorana or sterile neutrinos. One more important  aspect of this paper is that the Model-II predicts experimentally accessible neutrinoless double beta
($0\nu\beta\beta$) decay rate close to the current experimental limits
irrespective of light neutrino mass hierarchy. 
  
 For the gauge kinetic field tensor we have
\begin{eqnarray}
  F_{\mu\nu}&=&\partial_{\mu}W_{\nu}-\partial_{\nu}W_{\mu}+ig[W_{\mu},W_{\nu}],\nonumber\\
  W_{\mu}&=&\frac{1}{4}\sum_{i,j=1}^{10}\sigma^{ij}W_{\mu}^{ij},C=-\frac{\kappa}{8}\label{def}
 \end{eqnarray}
where $\frac{1}{2}\sigma^{ij}(W_{\mu}^{ij})$ ,and i,j=1,2,3......10
denote the 45 generators (gauge bosons)  of $SO(10)$.
 The GUT-scale  boundary conditions are modified by the 
 ${\rm dim}.5$ operator
 \begin{eqnarray}
 &&\alpha_{2L}(M_{U})(1+\epsilon_{2L})=\alpha_{2R}(M_{U})(1+\epsilon_{2R})=
\nonumber\\
 &&\alpha_{BL}(M_{U})(1+\epsilon_{BL})=\alpha_{3C}(M_{U})(1+\epsilon_{3C})=\alpha_{G},
\label{bc}
\end{eqnarray}
where the $\epsilon_i$ terms arise due to the  ${\rm dim}.5$ operator
 and $\alpha_{G}$ is the effective GUT fine structure constant which is predicted in terms of RG coefficients and the $\epsilon_i$ parameters.
The resulting analytic formulas for the unification mass $M_U$
, the  LR scale $M_R$  and GUT fine structure constant $\alpha_{G}$  are 
\cite{mkpjcp:1984,mkp:1987,ppdc}

\begin{align}
\ln\frac{M_{R}}{M_{Z}} & =\frac{1}{(XZ^{\prime}-X^{\prime}Z)}[(XP_{s}-X^{\prime}P_{\theta})+(X^{\prime}\rho_{2}-X\Sigma_{2})
 -\frac{2\pi}{\alpha_{G}}(X\epsilon^{\prime\prime}-X^{\prime}\epsilon^{\prime})],\label{req}\\ 
 \ln\frac{M_{U}}{M_{Z}} & =\frac{1}{(XZ^{\prime}-X^{\prime}Z)}[(Z^{\prime}P_{\theta}-ZP_{s})+(Z\Sigma_{2}-Z^{\prime}\rho_{2})
 -\frac{2\pi}{\alpha_{G}}(Z^{\prime}\epsilon^{\prime}-Z\epsilon^{\prime\prime})],\label{ueq} \\
 \frac{1}{\alpha_G}& = \frac{1}{D}\bigg [\frac{a_{3c}^{\prime}}{\alpha(M_z)}-\frac{a_{2L}^{\prime}+a_{2R}^{\prime}+\frac{2}{3}a_{BL}^{\prime}}{\alpha_S(M_z)}
 \nonumber \\
&  +\frac{1}{2\pi}\left(a_{3c}(a_{2L}^{\prime}+a_{2R}^{\prime}+ \frac{2}{3}a_{BL}^{\prime})-a_{3c}^{\prime}(\frac{5}{3}a_y+a_{2L})\right)\nonumber \\
&  \times \left(\frac{X(P_s-\Sigma_2)+X^{\prime}(\rho_2-P_{\theta})}{XZ^{\prime}-X^{\prime}Z}\right)\bigg ],\label{ageq}
\end{align}

The terms on the RHS reduce to the usual two-loop RG equations
in the limit $\epsilon^{\prime}=\epsilon^{\prime\prime}=\epsilon=0$.\\
It is well known that at one-loop level in such cases the effect of the GUT coupling cancels out from the combinations  ${\alpha(M_{Z})}^{-1}\left[1 -(8/3)\sin^2\theta_{W}(M_Z)\right ]$ and $\frac{1}{\alpha(M_{Z})}-(8/3)\frac{1}{\alpha_S(M_Z)}$
without affecting precise predictions of $M_U$ and $M_R$. Also the GUT
coupling is exactly determined in terms of one-loop and two-loop coefficients, and the value of
 ${\alpha(M_{Z})}^{-1}$  as can be seen by RG evolution of the latter.  But in the presence of 
the ${\rm dim}.5$ operator quite significant corrections arise because of smallness
of $\alpha_G$ as is evident from the third terms in the RHS of
eq.(\ref{req}) and  eq.(\ref{ueq}). Similarly RGE for fine structure
constant gives quite significant corrections inversely proportional to
$\alpha_G$ tending to smear out its precise value by Planck-scale
effect or the gravitational effect.
This is prevented by fixing the value of the GUT coupling by
eq.(\ref{ageq}) which is the RGE for $\alpha_{M_Z}$ with $M_U$ and
$M_R$ eliminated using eq.(\ref{req}) and  eq.(\ref{ueq}).
 We note that we have  four equations,  
 eq.(\ref{req}), eq.(\ref{ueq}),  eq.(\ref{ageq}), and
 eq.(\ref{espeq})(noted below) for four unknowns
 $M_R$,$M_U$, $\alpha_G$, and $C$, respectively.
Various symbols occurring in eq.(\ref{req}), eq.(\ref{ueq}), and eq.(\ref{ageq}) are
 \begin{align}
  P_{s} & =\frac{2\pi}{\alpha(M_{z})}\left(1-\frac{8}{3} \frac{\alpha(M_{z})}{\alpha_{s}(M_{z})}\right)\nonumber, \\
  P_{\theta} & =\frac{2\pi}{\alpha(M_{z})}\left(1-\frac{8}{3} \sin^2{\theta_{W}(M_{z})}\right)\nonumber,\\
  X & = a_{2R}^{\prime}+\frac{2}{3}a_{BL}^{\prime}-\frac{5}{3}a_{2L}^{\prime}\nonumber,\\
  Z & =\frac{5}{3}(a_{Y}-a_{2L})-\left(a_{2R}^{\prime}+\frac{2}{3}a_{BL}^{\prime}-\frac{5}{3}a_{2L}^{\prime}\right)\nonumber ,\\
  X^{\prime} & =a_{2R}^{\prime}+\frac{2}{3}a_{BL}^{\prime}+a_{2L}^{\prime}-\frac{8}{3}a_{3C}^{\prime}\nonumber,\\
   Z^{\prime} & =\frac{5}{3}a_{Y}+a_{2L}-\frac{8}{3}a_{3C}-\left(a_{2R}^{\prime}+\frac{2}{3}a_{BL}^{\prime}+a_{2L}^{\prime}-\frac{8}
   {3}a_{3C}^{\prime}\right)\nonumber,\\
  \rho_{2} & =2\pi\left[\theta_{2R}^{\prime}+\frac{2}{3}\theta_{BL}^{\prime}-\frac{5}{3}\theta_{2L}^{\prime}+\frac{5}{3}
  (\theta_{Y}-\theta_{2L})\right] \nonumber ,\\
   \Sigma_{2} & =2\pi\left[\theta_{2R}^{\prime}+\frac{2}{3}\theta_{BL}^{\prime}+\theta_{2L}^{\prime}-\frac{8}{3}\theta_{3C}^{\prime}+
   \frac{5}{3}\theta_{Y}+\theta_{2L}-\frac{8}{3}\theta_{3C}\right] \nonumber,\\
   \epsilon^{\prime} & =\epsilon_{2R}+\frac{2}{3}\epsilon_{BL}-\frac{5}{3}\epsilon_{2L}\nonumber,\\
   \epsilon^{\prime\prime} & =\epsilon_{2L}+\epsilon_{2R}+\frac{2}{3}\epsilon_{BL}-\frac{8}{3}\epsilon_{3C},\nonumber\\
   D & =a_{3c}^{\prime}\left(\frac{8}{3} +\epsilon_{2L}+\epsilon_{2R}+\frac{2}{3}\epsilon_{BL}\right)-(1+\epsilon_{3C})
   \left(a_{2L}^{\prime}+a_{2R}^{\prime}+ \frac{2}{3}a_{BL}^{\prime}\right) \nonumber\\
  &  +\frac{X\epsilon^{\prime\prime}-X^{\prime}\epsilon^{\prime}}{XZ^{\prime}-X^{\prime}Z} 
    \left[a_{3c}(a_{2L}^{\prime}+a_{2R}^{\prime}+ \frac{2}{3}a_{BL}^{\prime})-a_{3c}^{\prime}(\frac{5}{3}a_y+a_{2L})\right]\nonumber,\\
    \theta_i &=\frac{1}{4\pi}\sum_j\frac{b_{ij}}{a_j}\ln\frac{\alpha_j(M_R)}{\alpha_j(M_Z)}\nonumber,\\
    \theta_i^{\prime} &=\frac{1}{4\pi}\sum_j\frac{b_{ij}^{\prime}}{a_j^{\prime}}\ln\frac{\alpha_j(M_U)}{\alpha_j(M_R)}.\label{epp}
\end{align}
 The first, second, and third terms in the R.H.S. of eq.(\ref{req}) and eq.(\ref{ueq}) represent one loop, two-loop and gravitational effects respectively.
 In particular the combined GUT symmetry breaking VEVs of $ <\eta(1,
 1, 1)>, <\eta^{\prime}(1, 1, 15)> \subset 210_{H}$ can be written as
 \cite{pp}
 \begin{eqnarray}
\langle{\phi_{(210)}}\rangle= \frac{\phi_{0}}{8\sqrt{2}} \left(- \Gamma_{1}\Gamma_{2}\Gamma_{3}\Gamma_{4}+\Gamma_{1}\Gamma_{2}\Gamma_{5}\Gamma_{6}+
\Gamma_{3}\Gamma_{4}\Gamma_{5}\Gamma_{6} +
\Gamma_{7}\Gamma_{8}\Gamma_{9}\Gamma_{10}\right), \label{vev210} 
\end{eqnarray}
 where we have used\\  
$\Phi_{(210)}=\frac{1}{4!}\Gamma_{i}\Gamma_{j}\Gamma_{k}\Gamma_{l}. \phi^{ijkl}$,\\
and $<\Phi^{1234}>=<\Phi^{1256}>=<\Phi^{3456}>=<\Phi^{78910}>$~ leading to
  \begin{eqnarray}
  \epsilon_{2R}=-\epsilon_{2L}=-\epsilon_{3C}=\frac{1}{2}\epsilon_{BL}=\epsilon,  \nonumber \\
  \epsilon= -\frac{C M_{U}}{2
    M_{Planck}}\Bigl(\frac{3}{2\pi\alpha_{G}}\Bigr)^{\frac{1}{2}}.\label{espeq}
\end{eqnarray}

 With $\epsilon$ as input the GUT coupling $\alpha_G$ is at first
 determined using eq.(\ref{ageq}). The mass scales $M_U$ and $M_R$ are
 then determined from  eq.(\ref{ueq}) and eq.(\ref{req}).
Finally eq.(\ref{espeq}) determines $C$ as all other quantities in this
relation have been thus determined.
  Thus the single extra parameter $C$ as the coefficient of the ${\rm dim}.5$ operator in each model 
  brings down the $M_R$ to the LHC scale. It is to be noted that although $\alpha_G$ is small, the smallness of $\alpha_G$ alone
can not ensure TeV scale RH gauge bosons in the LRS model as discussed in ref.\cite{pp}. The low-mass $W_R^\pm$ bosons are favoured in the parity violating
 LR model with $g_{2L}\neq g_{2R}$, the asymmetry being generated by
 gravity induced ${\rm dim}.5$ operator with $\epsilon_{2L}=-
 \epsilon_{2R}=-\epsilon$. 

The light Higgs content that determines the
 one and two-loop coefficients depends upon whether LHC confirms
  trilepton production signals or dilepton production signals along
  with dominant double beta decay rate by ongoing experiments in the
  latter case. This gives rise to two different cases, Model-I and
  Model-II, as discussed below.

\subsection{\bf Two models with extra fermion singlets}
We now attempt to address the issue of neutrino masses and
mixings in the context of such TeV scale  $G_{2213}$ model descending
from non-SUSY $SO(10)$.
An additional advantage of doing GUT embedding through SO(10) is its
ability to  fit all charged
fermion masses and mixings
while explaining neutrino oscillation data through see saw mechanisms.
  As noted in Sec.1, because of the SO(10) constraint, $M_D\sim M_u$,
the type-I seesaw in SO(10) at the TeV seesaw scale fails to explain
the neutrino oscillation data.
 An interesting resolution of this problem can be made by using TeV
 scale inverse seesaw formula which has been suggested since 1986
 \cite{rnmvalle} and  extensively applied in the fermionic
extensions of the SM, LR gauge theory, and in SUSY SO(10)
\cite{psb-rnm:2010,blanchet-psb-rnm:2010} or non-SUSY SO(10) \cite{app,pas,psb-rnm:2015,ap}
where both the RH neutrinos $N_i$ and the extra fermion singlets $S_i$
collaborate to implement the mechanism. In
the SM, in addition to three RH neutrinos, $N_i (i=1,2,3)$, three
extra fermion singlets, $S_i (i=1,2,3)$,  are needed to implement the
inverse seesaw at the TeV scale.  
In the LR models, where RH neutrinos are already present as
fundamental representations, three additional singlet fermions $S_i(i=1,2,3)$
are added  for achieving inverse seesaw. In SO(10) where the RH neutrinos are
in the spinorial representation $ {16}_i$,  three
additional fermion singlets $S_i (i=1,2,3)$ are needed to implement the mechanism. 
In these cases instead of breaking the LR gauge
theory by ${126}_H$ of $SO(10)$, the original proposal in the minimal
 inverse seesaw
model is implemented through  the VEV of the RH doublet in ${16}_H
\subset SO(10)$ which also 
generates the $N-S$ mixing mass $M$ that occurs in the inverse seesaw
formula of eq.(\ref{inv}). This is discussed below under Model-I. Another verifiable
 prediction of  TeV scale inverse seesaw mechanism is the leptonic
nonunitarity effect detectable at long-baseline neutrino oscillation
experiments \cite{Malinsky:nonU} which is otherwise negligible in the
SM. Whereas the SM has negligible predictions for branching
ratios for charged lepton flavour violating (LFV) decays such as  $\mu \to
e\gamma$, $\tau \to e\gamma$,  and  $\tau \to \mu\gamma$, the inverse
seesaw mechanism predicts them only about few to four orders less than
their current experimental limits. Such rich structure of physical phenomena
realised within the inverse seesaw mechanism emhasizes the need of
extra non-standard fermion singlets $S_i(i=1,2,3)$ of SO(10) into the
theory.\footnote{ Alternatively, these $G_{2213}-$ singlet
  fermions may belong to non-standard fermion representations
  ${45}_F\subset SO(10)$ \cite{mkprad:2011}. If all $G_{2213}$
  non-singlet fermions are degenerate near the GUT scale, there
  contributions would not affect the threshold contributions to the mass scale predictions carried
  out in this work, Although in $E_6$ theory
  each singlet of a fermion generation is part of its fundamental representation which decomposes
  under $SO(10)$ as $27 = 16 + 10 + 1$, in such a case one goes beyond the SO(10)
  frame work. }. The basic reason that permits the inverse
seesaw  to be operative at the TeV scale in the presence of singlet fermions is the occurence of the
small coefficient of the mass term $\mu_S.SS$ in the
corresponding Yukawa Lagrangian in the neutrino mass  formula of
eq.(\ref{inv}).
In the context of the SM extension with added $N_i$ and $S_i$ ($i=1,2,3$), all neutrino
masses predicted by the inverse seesaw  vanish as
$\mu_S\to 0$ and the global lepton number symmetry is restored. This
phenomenon predicts $\mu_S$ to be a naturally small in the 't Hooft
sense \cite{tHooft} that plays a crucial role in bringing down the seesaw scale to
$M\sim {\cal O}(1)$ TeV, even when $M_D\sim M_u$. Numerous applications
of this formula are available with profound
new physics predictions in SM extensions \cite{inv:appl}, non-SUSY SO(10) with
low-mass $Z'$ boson \cite{ap}, SUSY SO(10) with TeV scale $G_{2213}$ symmetry and heavy pseudo-Dirac neutrinos mediating non-unitarity effects, LFV  decays  \cite{psb-rnm:2010}, and leptogensis 
\cite{blanchet-psb-rnm:2010}. In another approach in the extended seesaw frame
work of the SM \cite{Ellis:1993,grimus:2000,kk:2007,mmgsfv:2012} and in
 SUSY SO(10)\cite{skm:2007,mkparc:2010} heavy RH neutrino mass has been
introduced into the Yukawa Lagrangian and the neutral fermion mass matrix through the
intermediate scale value of  $<\Delta^0_R(1,3,-2,1)>= V_R\subset {126}_H\subset SO(10)$
\cite{app,pas,skm:2007,mkparc:2010,mkpsp:2012}.
The generalised form of Yukawa Lagrangian at TeV scale after
decoupling of LH scalar fields is\\
\begin{eqnarray}
{\cal L}&=& Y^l {\overline \psi}_L \psi_R {\bf \Phi}+f \psi^c_R\psi_R
\Delta_R + Y_{\chi}{\overline \psi}_R S \chi_R \nonumber\\
&&+S^T\mu_SS + h.c.\label{Yuklag}
\end{eqnarray}
where $\psi_L(\psi_R)=$LH (RH) doublet leptonic representations $\subset {16}\subset SO(10)$
, ${\bf \Phi} (2, 2, 0,1)=$ bidoublet Higgs scalar $\subset {10}_H\subset SO(10)$, $\Delta_R(1,3,-2,1)=$ RH triplet Higgs scalar $\subset {126}_H\subset SO(10)$, and $\chi_R(1,2,-1,1)=$ RH doublet Higgs scalar $\subset {16}_H \subset SO(10)$. 

After assigning VEV to the respectivde Higgs fields leads   
 to the $9\times 9$ neutral fermion mass matrix in the $(\nu, N, S)$ basis
\begin{eqnarray}
\mathcal{M}  = \left(
\begin{array}{ccc}
0 & M_{D} & 0 \\
M_{D}^{T} & {\tilde M}_{N}  & M \\
0 & M^{T} & \mu_{S} \label{matnn}
\end{array}
\right) 
\end{eqnarray}
where ${\tilde M}_{N}=0(fV_R)$ in the absence (presence) of ${126}_H$
  in Model-I (Model-II) as discussed below and $M=Y_{\chi}<\chi_R>$ . Block diagonalisation of this matrix in both the
models has been shown \cite{ app,pas,kk:2007,skm:2007,mkparc:2010,mkpsp:2012} to lead to the inverse seesaw formula \cite{rnmvalle}
for light neutrino mass matrix
\begin{equation}
 m_{\nu}=\frac{M_{D}}{M}\mu_{S}{\Bigl(\frac{M_{D}}{M}\Bigr)}^{T}, \label{inv}
\end{equation}
where the derivation in Model-II has been carried out in the limit
\begin{equation}
|{\tilde M}_N| > |M| >> |M_D|, |\mu_S|,  \label{limit}
\end{equation}
 leading to the cancellation of  the  type-I seesaw contribution. 
 In the non-SUSY SO(10) and Pati-Salam model, the extended
seesaw structure has been generated with low mass $W_R,Z_R$ bosons to
predict new dominant contribution to double beta decay mediated by the light 
sterile neutrino of first generation  \cite{app,pas,mkpsp:2012}. 
While in all the above cases the active neutrino mass formula is the same as
the original proposal \cite{rnmvalle},                                  leptonic
non-unitarity 
effects, observable LFV decays, dominant 
double beta decay, and resonant leptogenesis mediated by sterile neutrinos 
have been implemented in the
presence of type-II seesaw dominated neutrino mass formula and TeV
scale $Z'$ boson in non-SUSY SO(10) in ref.\cite{bpnmkp:2015}. The light singlet sterile fermions in
these SO(10) models also mediate like-sign dilepton production
via displaced vertices in the presence of TeV scale $G_{2113}$
symmetry, $Z'$ boson, and RH neutrinos \cite{bpnmkp:2015}. 
More recently in the context of non-SUSY SO(10) with additional
scalars and fermions at the TeV scale and externally imposed discrete
$Z_2$ symmetry, a rich structure for neutrino physics has been shown
to emerge through eq.(\ref{matnn}) with attractive and unified 
explanations for like-sign dilepton events in $pp\to eejj$, diboson
and dijet resonances at the LHC along with dark matter. It has been
particularly emphasized that the generalized parameter space spanned
by  eq. (\ref{matnn}) is very effective in accounting for the ratio of like-sign to opposite sign dilepton production cross sections recently observed at the LHC \cite{psb-rnm:2015}.\\ 

In the present work, we find that the Planck-scale effects and RG
constraints in the minimal chain favors the following two classes of
models which also succeed in explaining the neutrino oscillation
data. The two models differ in predicting the nature of the heavy
neutrinos: pseudo-Dirac (Model-I) or Majorana (Model-II) leading to
two different signals at LHC as discussed below.

\par\noindent{\bf (a). Model-I: Heavy pseudo-Dirac neutrinos:-}\\

In this case ${210}_H$ breaks $SO(10)$ and $D-{\rm Parity}$ to
$G_{2213}$  which further breaks to SM by the RH Higgs  
doublet  $\chi_R (1,2,-1,1) \subset {16}_H$. The SM theory
breaks to the low-energy symmetry by the standard Higgs
doublet $h (2,1,1)\subset \Phi(2,2,0,1) \subset {10}_H$. With such
minimal Higgs content respective  beta function
coefficients are presented in Table \ref{tab:loop}. Three additional
singlet fermions ($S_i, i=1,2,3$), one for each generation  are added
in case of $SO(10)$ theory as explained above. In the absence of
${126}_H\supset \Delta_R (1,3,-2,1)$ this model gives the neutral
fermion mass matrix of eq.(\ref{matnn}) with ${\tilde M}_N=0$ at the
renormalizable level of Yukawa interaction  
although ${\rm dim.}5$ operator gives $M_N \le 0.1 $ eV which is
negligible compared to $|\mu_S|$ needed to fit the neutrino
oscillation data through the inverse seesaw formula of
eq.(\ref{inv})\cite{ap}.
The $N-S$ mixing mass in this model occuring in  eq.(\ref{matnn}) is 
$M=Y_{\chi}<\chi^0_R>$. The Model-I applications to explain the neutrino oscillation data,
prediction of LFV decays and trilepton production signals at LHC have
been discussed in  Sec.3, and Sec.5.

For this model the one and two-loop beta function coefficients are
shown in Table \ref{tab:loop}.
 Using the input values  $\sin^2\theta_{W}(M_{Z})=0.23126\pm 0.00005$,  $\alpha({M_{Z})}=1/127.9$ and 
$\alpha_{3C}(M_{Z})=0.1187\pm 0.0017$\cite{PDG:2014} in eq.(\ref{req}), eq.(\ref{ueq}), we obtain solutions for  
 $M_{R}$, $M_{U}$, and $\alpha_G$ as shown in Table \ref{tab:scale1} . The value of $g_{2R}(M_{W_R})$ is obtained by running down $g_{2R}(\mu),g_{(B-L)}(\mu)$ from $\mu=M_U$ to $\mu= M_{W_R}$ and by ensuring the matching condition $g^{-2}_Y=(3/5)g^{-2}_{2R}+(2/5)g^{-2}_{(B-L)}$ at  $\mu= M_{W_R}$. The value of $\sigma$ is hence determined for each $\epsilon$. With all
other quantities occuring in eq.(\ref{espeq}) being thus determined, it gives the value of the coefficient $C$ of the ${\rm dim}.5$ operator. These solutions
are presented in       
Table \ref{tab:scale1} except threshold effects which have been
discussed below separately.
 It is clear that the Planck-scale induced solutions  as low  as $V_{\chi_R}
\sim M_R \simeq 20 $ TeV are allowed
 and the model predicts the $W_R$ mass $M_{W_R} \simeq
g_{2R} V_{\chi_R} \simeq 10$ TeV in the case of
minimal combination of the light Higgs sector with only two doublets,
$D_{\phi} =D_{\chi} =1$. These RH mass scales are spread over the range $\sim {\cal O} (1-100)$ TeV by  threshold effects as noted below.

\begin{table}[h!]
\caption{one loop and two loop beta function coefficients for RG evolution of gauge couplings}
\begin{center} 
\begin{tabular}{| p{2.4cm} | p{2.1cm} | p{2.5cm} | p{4cm} |p{4cm}|p{4cm}|}
\hline Gauge symmetry & Higgs content &~~~~~~${\bf{a_{i}}}$ &~~~~~~~${\bf{b_{ij}}}$\\ \hline 
 $G_{2213}$(Model:I), $D_{\phi}=D_{\chi}=1$ & $\phi(2,2,0,1)$, $\chi_{R}(1,2,-\frac{1}{2},1)$ & 
 \centering
 $\begin{pmatrix}                    
 -3 \\ \frac{-17}{6} \\ \frac{17}{4} \\ -7
 \end{pmatrix}$ 
 & 
 $\begin{pmatrix} 
    8 & 3 & \frac{3}{2}& 12 \\
    3 & \frac{61}{6} & \frac{9}{4}& 12             \\
    \frac{9}{2} & \frac{27}{4} &\frac{37}{8}& 4 \\
    \frac{9}{2} & \frac{9}{2} &\frac{1}{2}& -26
   \end{pmatrix}$  
   \\ \hline 
 $G_{2213}$(Model:II), $D_{\phi}=D_{\chi}=T_{\Delta_R}=1$ & $\phi(2,2,0,1)$, $\chi_{R}(1,2,-\frac{1}{2},1)$, $\Delta_R(1,3,-1,1)$ & 
  \centering
 $\begin{pmatrix}                    
 -3 \\ \frac{-13}{6} \\ \frac{23}{4} \\ -7
 \end{pmatrix}$      
 &   
 $\begin{pmatrix} 
    8 & 3 & \frac{3}{2}& 12 \\
    3 & \frac{173}{6} & \frac{57}{4}& 12             \\
    \frac{9}{2} & \frac{171}{4} &\frac{253}{8}& 4 \\
    \frac{9}{2} & \frac{9}{2} &\frac{1}{2}& -26
   \end{pmatrix}$  
 \\ \hline   
\end{tabular} 
\end{center}
\label{tab:loop}
\end{table}
     
\begin{table}[h!] 
\caption {Predictions for $M_R$, $M_U$, and the
  coupling constant ratio $\sigma ={g_{2L}^2\over g_{2R}^2}$ in 
  two $SO(10)$ models at two loop level including Planck-scale induced
  corrections.}
\centering 
\begin{tabular}{c c c c c c c c c  } 
\hline\hline 
  Model  & $\epsilon$ & $\alpha_{G}^{-1}$ & $M_{R}$(GeV) & $M_{U}$(GeV)& $\sigma=\left(\frac{g_{2L}}{g_{2R}}\right)^{2}$ &  $C$\\[0.5ex] 
\hline 
 Model-I & $0.086$ & $49.52$ & $ 2.49\times10^4$ & $9.59\times10^{15}$ & $1.288 $ & $-8.84$  \\ 
   & $0.087$ & $49.56$ & $ 2.1\times10^4$ & $9.51\times10^{15}$ &
 $1.291 $ & $-9.02$ \\
\hline 
  Model-II & $0.048$ & $47.69$ & $ 3.36\times10^4$ & $9.74\times10^{15}$ & $1.238 $  & $-4.95$  \\ 
   & $0.049$ & $47.72$ & $ 2.73\times10^4$ & $9.64\times10^{15}$ & $1.241 $ & $-5.1$ \\ 
  & $0.05$ & $47.77$ & $ 2.21\times10^4$ & $9.55\times10^{15}$ & $1.244 $ & $-5.2$ \\  [1ex] 
\hline 
\end{tabular} 
\label{tab:scale1} 
\end{table} 

\par\noindent{\bf (b).Model-II: Heavy Majorana neutrinos}\\ 
 
 For this purpose, in
addition to the Higgs representations of Model-I, we require the
$SO(10)$ representation ${126}_H \supset \Delta_{R} (1,3,-2,1)$ under
$G_{2213}$ that carries $B-L= -2$ with corresponding  
 coefficients given in Table \ref{tab:loop}.
 When the RH triplet  acquires VEV $\left<\Delta_{R}^{0}\right> =V_{\Delta_R}$,  $G_{2213}$ symmetry is broken down to SM  and RH
 neutrinos acquire heavy masses through Yukawa interaction term $f 16. 16.
 {126}_H$ leading to ${\tilde M}_N= fV_{\Delta_R}$ that replaces the central
 part of the $3\times 3$ null matrix of eq.(\ref{matnn}). The RH doublet
 $\chi_R(1,2,-1 ,1) \subset {16}_H$, apart from taking part in  
symmetry breaking process rather weakly, generates the N-S mixing mass term $M$ as
noted in the case of Model-I leading to gauged inverse seesaw formula for
neutrino masses provided ${\tilde M}_N > M > M_D, \mu_S$, a condition well
known in extended seesaw mechanism \cite{grimus:2000,kk:2007}.
The would-be dominant type-I seesaw term in this model cancels out in
such decoupling limit \cite{app,pas,kk:2007} leading to
gauged inverse seesaw formula of eq.(\ref{inv})  to explain the
neutrino oscillation data. There are two heavy Majorana neutrino mass
matrices: $m_N$ for RH neutrino and $m_s$ for sterile neutrino under
the constraint  $m_N >> m_s$ 
  \begin{equation}
  m_{s}=\mu_{s}-M\frac{1}{{\tilde M}_{N}}M^{T}+...\label{majms}
 \end{equation}
 The heavy RH Majorana neutrino mass matrix is very close to its
 gauged value,
 \begin{equation}
  M_{N}={\tilde M}_{N}+...
 \end{equation}
These two types of heavy Majorana neutrinos emerging as a result of
the gauged
extended seesaw mechanism can mediate neutrinoless double beta decay in
the $W_{L}-W_{L}$, $W_{L}-W_{R}$, and the $W_{R}-W_{R}$
channels. Further both of them are capable of mediating the dilepton
 production process at the LHC.\\
  
Using numerical values of $a_i$ and $b_{ij}$ in eq.(\ref{req}),
eq.(\ref{ueq}), and eq.(\ref{epp}) and following the same procedure as
outlined for Model-I, 
  the solutions for mass scales   $M_{R}$ and $M_{U}$, and $C$ and
  $\sigma$  are also presented for this Model-II  in  Table \ref{tab:scale1}. It is clear that
  in this case  low-mass RH gauge bosons $\sim 10$ TeV
 are  permitted at the
  LHC energy scale for $\sigma \simeq 1.24$  which may lead to the interpretation that the
  additional corrections could be due to quantum gravity effects.  
  It is interesting to note that $\sigma = {g_{2L}^2\over
  g_{2R}^2}\simeq 1.24-1.29$ is only $24\%-29\%$ larger compared to its
 value
$\sigma = 1$ in the manifest LRS model \cite{rnm}. These correspond to
 the $W_R$ gauge couplings $g_{2R}=0.56(0.57)$ in Model-I (Model-II)
 at $M_{W_R}=2 $ TeV.

\subsection{\bf GUT threshold effects}
  
As the representations $210_H$ or $126_H$ have a number of superheavy components around the GUT scale, we have estimated  their threshold effects
\cite{mkpcch,mp}
on $M_R$, $M_U$ and proton lifetime\cite{lmpr,mkpcch,mp}. Following the steps those led to eq.(\ref{req})-eq.(\ref{ageq}), the  analytic formulas for
threshold corrections for mass scales are

\begin{equation}
 \Delta\ln\frac{M_R}{M_Z}=\frac{X^{\prime}\rho_{\Delta}-X\Sigma_{\Delta}}{XZ^{\prime}-X^{\prime}Z} 
\end{equation}
\begin{equation}
 \Delta\ln\frac{M_U}{M_Z}=\frac{Z\Sigma_{\Delta}-Z^{\prime}\rho_{\Delta}}{XZ^{\prime}-X^{\prime}Z} 
\end{equation}
where
\begin{align}
 \rho_{\Delta} & =-2\pi\left[\Delta_{2R}^{\prime}+\frac{2}{3}\Delta_{BL}^{\prime}-\frac{5}{3}\Delta_{2L}^{\prime}\right],\nonumber\\
 \Sigma_{\Delta} & =-2\pi\left[\Delta_{2R}^{\prime}+\frac{2}{3}\Delta_{BL}^{\prime}+\Delta_{2L}^{\prime}-\frac{8}{3}\Delta_{3c}^{\prime}\right],\nonumber \\
 \Delta_{i}^{\prime} & =\sum_{\alpha}\frac{b_i^{\alpha}}{12\pi}\ln\frac{M_{\alpha}}{M_U},i=2L,2R,BL,3C,
\end{align}
 $M_{\alpha}$ being the superheavy component mass of
the Higgs representation. Assuming that all superheavy components of a
GUT representation have a common mass \cite{lmpr,mp}, the corrections are
 shown in Table \ref{tab:tu}, where the first(second) line gives maximized uncertainty in $M_U$($M_R$) in both models. It is clear that the lifetime 
prediction including GUT threshold effects can be accessible to ongoing searches\cite{sk-hk}.

\begin{table}[h!] 
\caption { Threshold effects on predicted mass scales and proton lifetime where the results given in the first(second) line in each model are due to
 maximization of uncertainty in $M_U$($M_R$). The factor $10^{\pm 0.2}$ arises due to $1\sigma$ uncertainty in $\sin^2{\theta_W}(M_Z)$ and $\alpha_S(M_Z)$.}
\centering 
\begin{tabular}{c c c c  } 
\hline\hline 
  Threshold Uncertainty  & $\frac{M_R}{M_{R^0}}$ & $\frac{M_U}{M_{U^0}}$ &$\tau_p$(yrs.) \ \\[0.1ex] 
\hline 
 Model-I & $10^{\pm 0.006}$ & $10^{\pm 0.364}$ & $10^{36.15 \pm 1.456\pm0.2}$   \\
   & $10^{\pm 0.332}$ & $10^{\pm 0.205}$ & $10^{36.15 \pm 0.82\pm0.2}$ \\
\hline 
  Model-II & $10^{\pm 0.76}$ & $10^{\pm 1.45}$ & $10^{36.15 \pm 5.8\pm0.2}$\\
   & $10^{\pm 1.548}$ & $10^{\pm 0.47}$ & $10^{36.15 \pm 1.88\pm0.2}$ \\ [1ex] 
\hline 
\end{tabular} 
\label{tab:tu} 
\end{table} 

Apart from these renormalisable threshold corrections, the other
possible corrections may be due to two more non-renormalisable 
 ${\rm dim}.6$ operators such as $Tr({F^{\mu\nu}\Phi_{210}^2F_{\mu\nu}})$,
$Tr({\Phi_{210}F^{\mu\nu}\Phi_{210}F_{\mu\nu}})$ which introduce two
more unknown parameters. Contributions of other Higgs fields to ${\rm
  dim}.6$ operators are negligible because of their smaller VEVs.
 In the spirit of
earlier approaches that quantum gravity effects are reflected most dominantly via
Planck-scale induced  ${\rm dim}.5$ operators, they are ignored in the
minimal model with minimal number of parameters. Even if they are included, we do not think these
contributions to be relevant because of the following:
(i) the correction to GUT-gauge coupling $\alpha_i$  due to operator of
${\rm dim}. n >4 $ is $C_{ n} {(\frac{M_U}{M_{\rm
      Planck}})}^{n-4}$. Even if the coefficient $C_{n}$
does not decrease with $n$, treating  $C_n \sim {\cal O}(1)$, the higher
order terms are reduced by ${\cal O}[10^{-3(n-4)}]$ which may be
considered negligible for $n \ge  6$.(ii) Even if we include these corrections
, no new interesting physics is expected to
emerge as we have already achieved LHC scale LR gauge theory and, of
course, experimentally observable proton decay by including threshold effects.
 
\section{Neutrino masses and lepton flavor violation}
The Dirac neutrino mass matrix $M_D$ occurring in eq.(\ref{inv}) is determined by the GUT-scale fitting of the
extrapolated values of all charged fermion masses
obtained by
following the bottom-up approach \cite{dp} and running it down 
to the TeV scale following top-down approach as explained in the
corresponding cases \cite{app,pas,ap}. While the procedure followed in
ref.\cite{ap} is used for Model-I, the procedures followed in
ref.\cite{app,pas} is utilized for Model-II. \footnote{ Any additional
  $SO(10)$ Higgs representations or higher dimensional operators which
  may be needed for charged fermion mass fits at the GUT scale do not affect the LHC scale particle spectrum.} An additional bidoublet
$\phi^{\prime}\subset {10}_{H'}$ is needed to fit fermion
masses without affecting coupling unification substantially.   
The Higgs bidoublet $\xi(2,2,15) \subset {126}_H$ acquires the induced
VEV $v_{\xi}\simeq 10-50$ MeV \cite{baburnm} which
,along with the direct VEVs of the two bidoublets, enables fitting 
 all charged fermion masses in Model-II. 
A byproduct of this fitting is the diagonalised version of the heavy
RH neutrino mass matrix ,
\begin{eqnarray}
{\hat M}_N &=&{\hat f}V_R \nonumber\\
            &&={\rm diag}. ({\hat M}_{N_1}, {\hat M}_{N_2}, {\hat
  M}_{N_3}).\label{mneq}
\end{eqnarray}
where, in our Model-II,
\begin{eqnarray}
{\hat M}_{N_1} & \simeq & 150 {\rm GeV}- 1.5 {\rm TeV},\nonumber\\
{\hat M}_{N_2} & \simeq & 500 {\rm GeV}- 3.0 {\rm TeV},\nonumber\\
{\hat M}_{N_3} & \simeq & 2.0 {\rm GeV}- 7.5 {\rm TeV}.\label{mnnum}   
\end{eqnarray}

In the absence of ${126}_H$
 in Model-I, the dim.6 operator
 ${F^a_{ij}16_i.16_j.{10}_{H_a}{45}_H{45}_H}/{M^{\prime}}^2$
 discharges the equivalent role where $M^{\prime}\sim
 M_{Planck}$.  This ${45}_H$ remains near the GUT scale without
 affecting the particle spectrum at the LHC scale.
Up to a good approximation,
in both the models the
 Dirac neutrino mass matrix at the LHC scale is
\begin{eqnarray}
 M_{D}(M_{R}^{0})  = \left(
\begin{array}{ccc}
0.0151 & 0.0674-0.0113i & 0.1030-0.2718i \\
0.0674+0.0113i & 0.4758  & 3.4410+0.0002i \\
0.1030+0.2718i & 3.4410-0.0002i & 83.450
\end{array}
\right)\ GeV. \label{MD}
\end{eqnarray}  
 
  The dominant
source of LFV is through the $W_L$-loop in both the
models and there are two types of heavy Majorana fermion
exchange contributions in case of Model-II. The RH neutrino exchange
contribution can be considered subdominant since $M_{N_i} >> M_i$.
Using the relevant analytic formulas \cite{pilaf}
  we estimate LFV decay
 branching ratios $\mu\rightarrow e+ \gamma$, $\tau\rightarrow e+ \gamma$ and
 $\tau\rightarrow \mu+ \gamma$ as shown in  Table.\ref{tab:lfvbr}
 where the allowed values of $M_i, (i=1,2,3)$ satisfying the
 non-unitarity constraints have been also given \cite{app,ap}.
  As the predicted  values are $3-5$ orders smaller than the current
 experimental limits, they may be  accessible to ongoing or planned
 searches with improved accuracy.
 \begin{table}[h] 
\caption {Nonunitarity predictions of branching ratios for charged lepton flavor violating decays
 as a function of pseudo Dirac neutrino masses.} 
\centering 
\begin{tabular}{c c c c c} 
\hline\hline 
 $M(GeV)$ & $\vert\hat{M}_S\vert$(GeV) & $BR(\mu \rightarrow e\gamma)$ & $BR(\tau \rightarrow e\gamma)$ & $BR(\tau \rightarrow \mu \gamma)$    \\ [0.5ex] 
\hline 
 $(50,200,1711.8)$ &$(10,50,837.21)$ &$1.19\times 10^{-16}$ & $4.13\times 10^{-15}$ & $5.45\times 10^{-13} $   \\ 
 $(100,100,1286)$  &$(12.5,20,661.5)$ & $1.07\times 10^{-15}$ & $2.22\times 10^{-14}$ & $2.64\times 10^{-12} $  \\ 
 $(100,200,1702.6)$  &$(16.6,40,828.24)$ & $1.14\times 10^{-16}$ & $4.13\times 10^{-15}$ & $5.52\times 10^{-13} $   \\ 
 [1ex]        
\hline 
\end{tabular} 
\label{tab:lfvbr} 
\end{table}  

Using a set of values on $M$, some of which are given in Table. \ref{tab:lfvbr}, and the
Dirac neutrino mass matrix from eq.(\ref{MD}), we fit
the available data on neutrino masses and mixings through inverse
seesaw formula of eq.(\ref{inv}) for all the three types of mass
hierarchies: NH, IH, and QD. A wide range of values of the matrix
elements of
$M={\rm diag.}(M_1, M_2, M_3)$
are allowed  consistent with LFV constraints and the neutrino
oscillation data \cite{nudata:1,nudata:2}. In each case the fit gives
a set of elements for $\mu_S$. Our solutions for the NH case indicated
by recent cosmological constraints \cite{cosmo:14,cosmo:15} is given
below for NH case  with ${\hat m}_{\nu}={\rm diag.}(0.001, 0.0088,
0.049)$ eV and $M={\rm diag.}(50, 200, 1712)$ GeV.   
\begin{eqnarray}
 \mu _{s}(\rm GeV)  = \left(
\begin{array}{ccc}
0.002+0.00001i & -0.0015-0.00001i & 0.0004-0.0002i \\
 -0.0015-0.00001i & 0.001 & -0.0003+0.0001i \\
0.0004-0.0002i &-0.0003+0.0001i& 0.00006-0.0001i
\end{array}
\right)\ GeV\label{musdet}  
\end{eqnarray}

Different aspects of LFV in non-SUSY $SO(10)$ have been discussed in
ref.\cite{app,pas,ap} and our predictions in the corresponding cases
are similar.

\section{Lepton number violation}
The standard contribution to neutrinoless double beta decay in the $W_{L}-W_{L}$ channel is due to light
neutrino exchanges. But because of the presence of mixing with the RH
neutrino and the extra fermion singlet states, the LH neutrino flavor
state $\nu_{\alpha L} (\alpha=e,\mu,\tau)$ is expressed in terms of the heavy and light mass eigen states
\begin{eqnarray}
\nu_{\alpha L}\sim {\cal V}^{\nu\nu}_{\alpha i}{\hat {\nu}}_i+{\cal
  V}^{\nu S}_{\alpha i}{\hat {S}}_i+{\cal V}^{\nu N}_{\alpha i}{\hat
  {N}}_i,\label{flmix}
\end{eqnarray}
where  ${\cal V}^{\nu\nu}_{e i}$ is approximated to be the standard PMNS
mixing matrix elements.
As already stated 
 ${\cal
  V}^{\nu S}_{e i}={(M_D/M)}_{ei}= {(M_D)}_{ei}/M_i$, and ${\cal V}^{\nu N}_{e i}={(M_D/M_N)}_{ei}$.
One important aspect of this Model-II is that even in the $W_{L}-W_{L}$ channel the singlet fermion exchange allowed within the extended seesaw mechanism 
can yield much more dominant contribution to $ 0\nu \beta  \beta$
decay rate with lifetime prediction close to the current experimental limits
\cite{bbexpt:1,bbexpt:2,bbexpt:3,bbexpt:4}. 
The contributions due to the exchanges of heavy $\Delta^{++}_L$,
$\Delta^{++}_R$, and  RH neutrinos  in the $W_R-W_R$ channel
\cite{Tello:2011}
 are  negligible 
 in this extended seesaw framework 
compared to those due to the light neutrino and the singlet sterile fermion exchanges
in the $W_L-W_L$ channel for which the three different contributions
to the
amplitude  and the corresponding mass parameters are
summarised in Table \ref{tab:lnvamp}.

\begin{figure}[h!]
\begin{center}  
\includegraphics[scale=0.6]{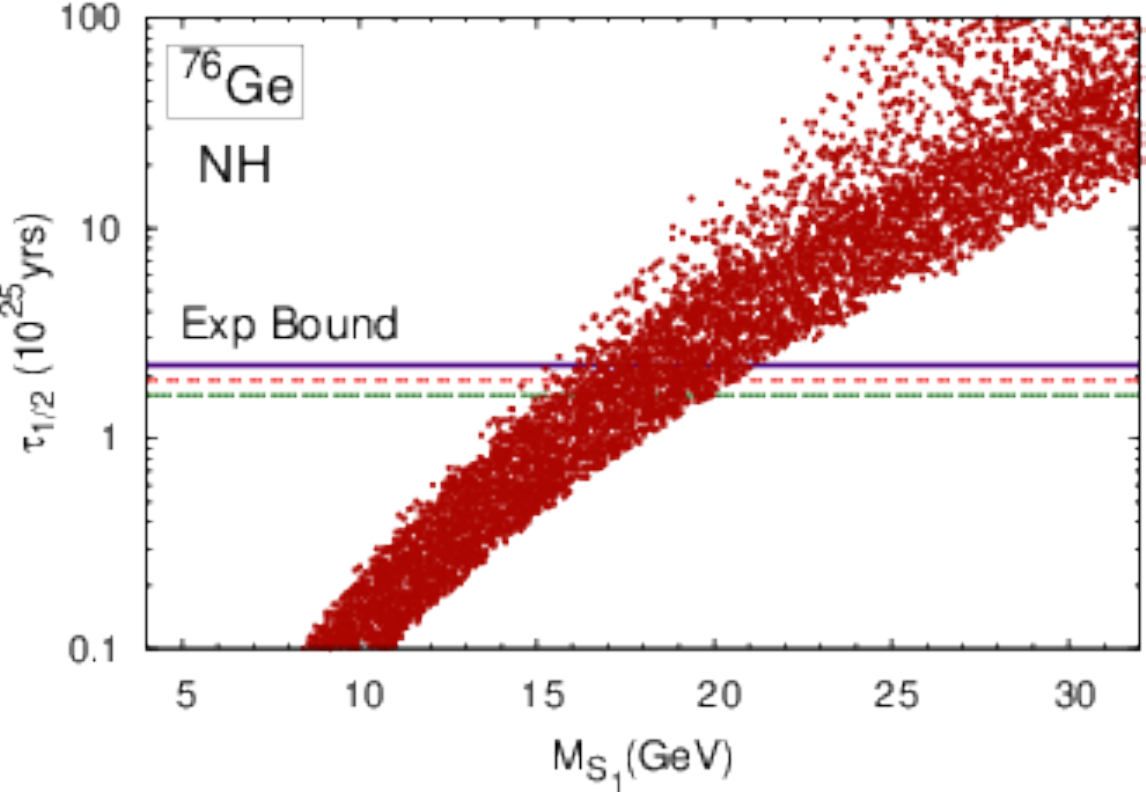}
\caption{Scattered plot for half-life of neutrino-less double beta
  decay as a function of $M_{S_1}$ in the case of NH light neutrino
  masses for $p=130-277$ MeV. The horizontal lines represent lower bounds on
  half-life obtained by different experimental groups. }
\end{center}  
\label{fig:scatNHGe}
\end{figure}

\begin{table}[h!] 
\caption {Formulas for amplitudes and effective mass parameters in
the $W_{L}-W_{L}$ channel for $0\nu\beta\beta$ decay where $|p|^2$ has
 been defined in the text.}
\centering 
\begin{tabular}{c c c c c  } 
\hline\hline 
  Channel  & Mediating particle & Amplitude & Effective mass parameter \ \\[0.1ex] 
\hline 
 $W_{L}-W_{L}$ & $\nu$  & $A_{\nu}^{LL}\propto \frac{g_{2L}^4}{M_{W_{L}}^{4}} \sum\limits_{i=1,2,3}\frac{(\mathcal{V}_{ei}^{\nu\nu})^{2}m_{\nu i}}
{p^{2}}$  &  $m_{\nu}^{ee,L}= \sum\limits_{i}\left(\mathcal{V}_{ei}^{\nu\nu}\right)^{2}m_{\nu_{i}}$   \\
       &$S$ & $A_{S}^{LL}\propto \frac{g_{2L}^4}{M_{W_{L}}^{4}} \sum\limits_{j=1,2,3}\frac{(\mathcal{V}_{ej}^{\nu S})^{2}}{m_{S_j}}$ &$m_{S}^{ee,L}= \sum\limits_{i}\left(\mathcal{V}_{ei}^{\nu S}\right)^{2}\frac{\left|p\right|^{2}}{m_{Si}}$  \\
  &$N$ &  $ A_{N}^{LL}\propto \frac{g_{2L}^4}{M_{W_{L}}^{4}} \sum_{k=1,2,3}\frac{(\mathcal{V}_{ek}^{\nu S})^{2}}{m_{N_k}}$  &  $m_{N}^{ee,L}= \sum\limits_{i}\left(\mathcal{V}_{ei}^{\nu N}\right)^{2}\frac{\left|p\right|^{2}}{m_{Ni}}$\\  [1ex] 
\hline 
\end{tabular} 
\label{tab:lnvamp} 
\end{table} 
 Since the sterile neutrino
mass eigen value ${\hat M}_{S_1}<< {\hat M}_{N_i}$ and the $N-S$
mixing elements can be made to satisfy $M_i<< M_{N_i}$ we obtain the
dominance of light sterile neutrino exchange contribution over the RH
neutrino exchange contribution in the $W_L-W_L$ channel since $|{
    m}^{ee,L}_{N}|<< |{ m}^{ee,L}_{S}|$.
Then using the mass parameters from Table \ref{tab:lnvamp},
the inverse half life can be written as
\begin{eqnarray}
  \left[T_{1/2}^{0\nu}\right]^{-1} &\simeq &
  G_{01}|\frac{{\cal M}^{0\nu}_\nu}{m_e}|^2|m_{eff}|^2\label{halflife}
\end{eqnarray}
where
\begin{eqnarray}
|m_{eff}|^2&=&|{m}^{ee,L}_{\nu} +{ m}^{ee,L}_{S}|^2,\label{meff1} 
\end{eqnarray}
where
\begin{eqnarray}
|m_{eff}|^2&=&|{m}^{ee,L}_{\nu}|^2 +|{ m}^{ee,L}_{S}|^2+{\rm I.T}. \label{meff2}.
\end{eqnarray} 
In eq.(\ref{meff2}) I.T.$=$ interference term between the two quantities  
$=2|{m}^{ee,L}_{\nu}||{m}^{ee,L}_{S}|\cos \gamma_{\nu S}$,
$\gamma_{\nu S}$ being their phase difference. Although it is possible
to adjust the phases of the two, especially those in ${m}^{ee,L}_{S}$,
resulting in  $\gamma_{\nu S}=(2n+1)\pi/2$ with $n={\rm integer}$ and
vanishing I.T., for numerical estimation of half-life we have taken the full
expression in eq.(\ref{meff1}). Details have been given in
ref.\cite{pas} where a new analytic formula for half-life has been
also reported.\\ 
In eq.(\ref{halflife})
 $G_{01}=$ phase space factor $=0.686\times 10^{-14}${\rm
  yrs}$^{-1}$, 
${\cal M}^{0\nu}_{\nu}=$ nuclear matrix element (NME)
  correspoding to light LH neutrino exchange, and $p$ denotes the
  neutrino virtuality momentum. In terms of ${\cal M}^{0\nu}_{\nu}$ and
    ${\cal M}_N^{0\nu}$, the NME corresponding to heavy neutrino exchanges
    , it is also expressed as \cite{Barry:2013,Doi:1985,Vergados:2002} 
$|p|^2= (m_pm_e)\frac {{\cal M}_N^{0\nu}}{{\cal
    M}_{\nu}^{0\nu}}$.
 Available values of NMEs with
uncertainties cover the range ${\cal M}_{\nu}^{0\nu}=2.58-6.64$,
${\cal M}_N^{0\nu}=232-242$ leading to  $|p|\simeq (130-277)$ MeV for
${}^{76}Ge$ isotope. 
Using eq.(\ref{halflife}) and eq.(\ref{meff1}), and Dirac and Majorana
phases, double beta decay half-life
 predictions have been discussed in detail showing saturation of
experimental limits for ${\hat M}_{S_1}=15-18$ GeV for three different
light neutrino mass hierarchies \cite{pas} where all possible
interference effects have been included for different active
neutrino mass hierarchies. It is interesting to note that in the case
of normally hierachical (NH) active neutrino masses, the lightest sterile neutrino contribution 
with mass ${\hat M}_{S_1}=5-40$ GeV dominates  the double beta
decay rate with $|m_{eff}|\simeq |{m}^{ee,L}_{S}|$.
Confining  to the
normally hierarchical (NH) light neutrino masses indicated by recent
cosmological bounds \cite{cosmo:14,cosmo:15}
\begin{eqnarray}
\sum_i {m}_{\nu_i} \le 0.12\,\, {\rm eV}, \label{cosbnd}
\end{eqnarray}
 and for naturally allowed values of ${\hat
  M}_{S_1} \sim 5-40$ GeV with ${\hat M}_{S_2}, {\hat M}_{S_3}>> {\hat
  M}_{S_1}$, the predictions in Model-II is given in
 Fig.\ref{fig:scatNHGe} for $p=130-277$ MeV, 
where the horizontal lines are the lower limits on the half-life measured
by different experimental groups.\cite{bbexpt:1,bbexpt:2,bbexpt:3,bbexpt:4}. 

Saturation
of current experimental bound on $0\nu\beta\beta$ decay half life gives the lower bound on the lightest
sterile neutrino mass, $m_{S_1} \ge 17 \pm 3$
GeV. Thus, the present TeV scale $G_{2213}$ model is found to be
capable of saturating the current experimental limits of neutrinoless
double beta decay in the $W_L-W_L$ channel where both the emitted
elctrons have left-handed chiralities and the lightest sterile neutrino
exchange dominates the process especillay for normally hierarchical
masses of light neutrinos as indicated by cosmological bounds.

\section{LHC signals of heavy neutrinos and $W_R$ boson}
 The Large Hardon Collider(LHC) offers an amazing opportunity to
 explore new
   physics beyond the electroweak scale . The LHC has already taken data at $\sqrt{s}=8$ TeV and expected to take 
further data at $\sqrt{s}=14$ TeV in run-II for physics signals beyond
the standard model. Very recently there have been various recent attempts to explain
observed excess of events beyond the standard model \cite{lhc}.  
Our model 
predicts $G_{2213}$ symmetry at lower scale  $\mu=M_R$ of the order of 1-10 TeV.  The $W_R$ bosons from   $M_R=1-10$ TeV can be clearly produced from pp
collision which can subsequently decay to a RH charged lepton and a RH neutrino. If the RH neutrino is Pseudo-Dirac, this will manifest into trilepton signals
or if it is a heavy Majorana neutrino, it can manifest into two like-sign dileptons and jets. In this section, we examine both the above possibilities. 

     At the LHC, the parton-level generation of a heavy neutrino can be realized in the following way
\begin{equation}
 q\bar{q}^{\prime}\longrightarrow W_{L}/W_{R}\longrightarrow l^{+}N(l^{-}\bar{N}),(l=e,\mu,\tau)
\end{equation}
provided this process is kinematically feasible. This has
lepton-number conserving (LNC) or lepton number violating (LNV) decay
modes depending on whether $N$  is pseudo 
Dirac as in Model-I or Majorana as in Model-II. 
We use the parton level differential cross section \cite{ddo}
 \begin{equation}
  \frac{d\hat{\sigma}_{LHC}}{d\cos\theta}=\frac{k\rho}{32\pi\hat{s}}\frac{\hat{s}+M^{2}}{\hat{s}}\frac{g^{4}}{48}
   \frac{(\hat{s}^{2}-M^{4})(2+\rho\cos^{2}\theta)}{(\hat{s}-m_{W}^{2})^{2}+m_{W}^{2}\Gamma_{W}^{2}}\label{dsigdc}
 \end{equation}
 where $k=3.89\times10^{8}$pb, $\hat{s}$ is the square of centre-of-mass energy of the colliding partons, $M$ is mass of $N$, and $\rho=
 {(\hat{s}-M^{2})}/{(\hat{s}+M^{2})}$.
 
 The total production cross section at the LHC is

\begin{equation}
 \sigma_{prod}= \frac{kg^4}{768\pi s}\int_{\tau_0}^1\gamma \frac{d\tau}{\tau} \int_\tau^1 \frac{dx}{x} 
 \left[f_{u}(x,Q) f_{\bar{d}}(\frac{\tau}{x},Q)+(u\rightarrow \bar{d},\bar{d}\rightarrow u)\right],\label{prodcross} 
\end{equation}
where $\tau=\hat{s}/{E_{CM}^{2}}$ and  $E_{CM}$ is centre-of-mass
energy of the LHC, and \\
$\gamma =((\hat{s}+M^{2})/\hat{s})\times((\hat{s}^{2}-M^{4})(2+\rho/3)/((\hat{s}-m_{W}^2)^2 + m_{W}^2 \Gamma_{W}^2)) $.

\begin{figure}[h!]
\centering
\subfigure[]{%
\includegraphics[scale=0.46]{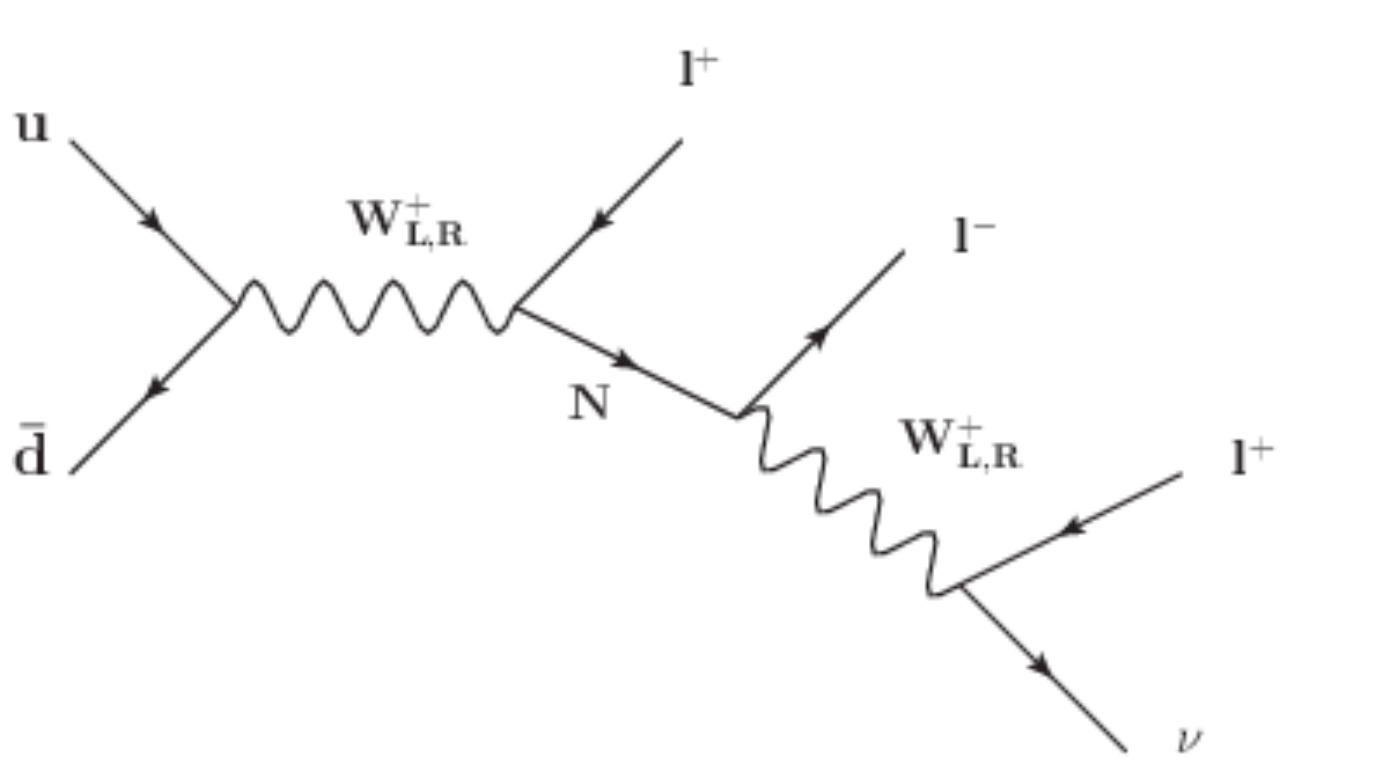}
\label{fig:trilep}}
\quad
\subfigure[]{%
\includegraphics[scale=0.46]{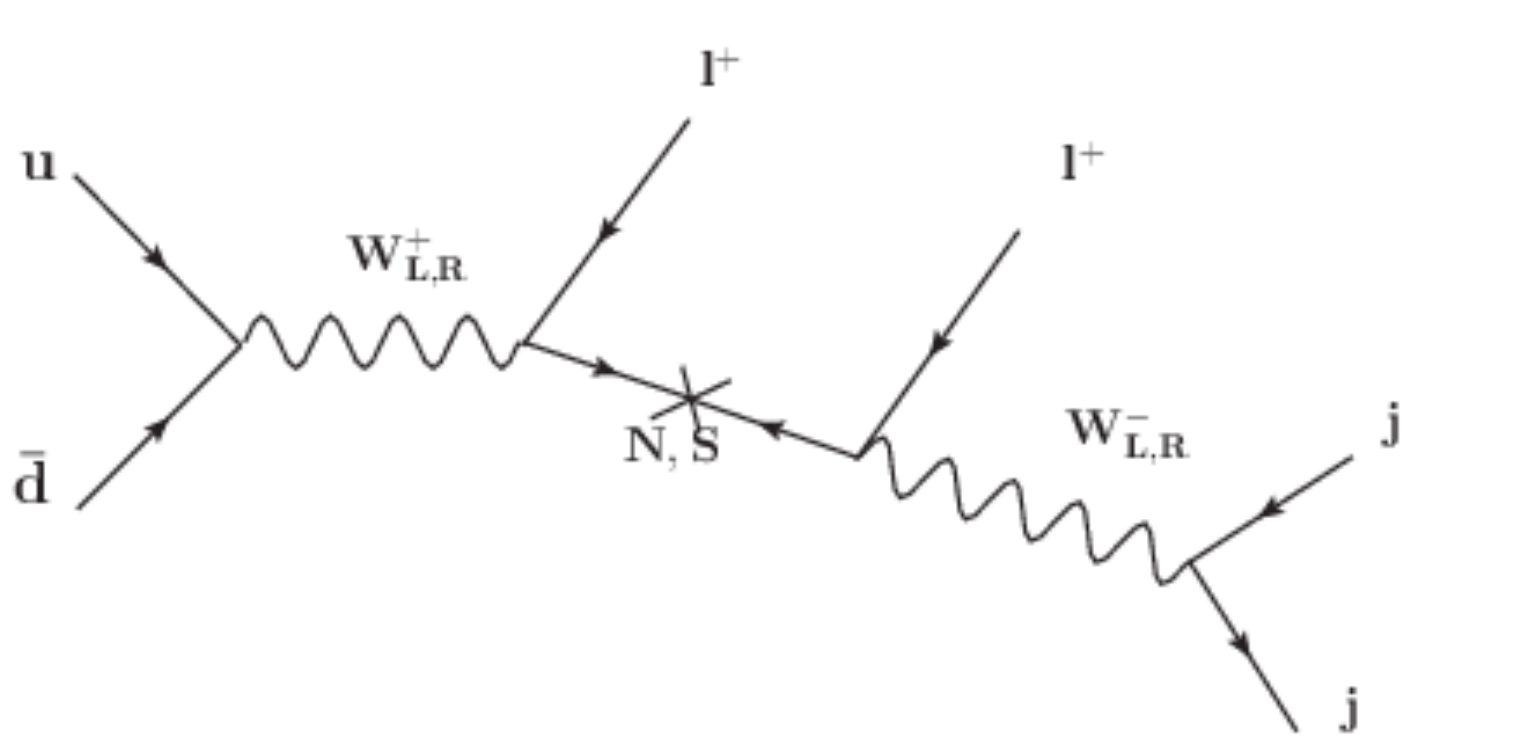}
\label{fig:dilep}}
\caption{ Feynman diagrams for trilepton (left-panel) and dilepton
  (right-panel) signals at the LHC in various channels: LL, RR, LR,
  and RL where instead of confining to the exchange of real $W_R$
  boson \cite{ks}, the general possibilities of including both real and
virtual $W_L,W_R$ exchanges \cite{cdm} in the second stage have been considered. }
\label{fig:tridi}
\end{figure}

The Feynman diagrams for trilepton(dilepton) production mechanism is
shown in the left-panel (right-panel) of Fig.\ref{fig:tridi}

\subsection{\bf Trilepton signals}

The RH neutrinos in Model-I being pseudo-Dirac neutrinos can not mediate
like-sign dilepton production. Also the opposite sign dilepton signal $l^\pm l^\mp jj$ is not a viable option as it is swamped with a large SM background.
The best channel for probing heavy pseudo-Dirac neutrinos is the trilepton mode  where $W_{L}/W_{R}$ decays to leptonic final states:
$pp \rightarrow W_{R}^{\pm}\rightarrow N \ell^{\pm} \rightarrow W_{L}/W_{R}^{\star}\ell^{\mp}\ell^{\pm}\rightarrow \nu\ell^{\pm}\ell^{\mp}\ell^{\pm}$
\cite{agu-saav}.

The inclusive cross-section for the trilepton state in a generic seesaw model is given by\cite{ddo}
\begin{equation}
 \sigma(pp\rightarrow
 l_{1}l_{2}l_{3}+T_{me})=\sigma_{prod}(pp\rightarrow
 W^{\star}\rightarrow Nl_{1})Br(N\rightarrow l_{2}W)Br(W\rightarrow
 l_{3}\nu).\label {inclsig}
\end{equation}
 Here,  $T_{me}$ stands for the missing transverse energy and the $W_{L}$ branching ratio $Br(W\rightarrow l\nu)$=0.21\cite{pdg}. 
 We have assumed $m_{N}> m_{W}$.  
Although this condition is needed for kinematic feasibility of the
decay $N\rightarrow l_2W \rightarrow l_2l_3\nu$
when the exchanged $W-$ boson is real, this is not required for virtual
$W^*$ exchange to give $N \rightarrow l_2W^* \rightarrow l_2l_3\nu$.
One important aspect of this model is that the fermion mass fitting
and LFV constraint predict all the elements of the heavy-light
neutrino mixing matrix $V_{\nu S}={M_D\over M}$. For example using
eq.(\ref{MD})
and $M_{N_2}\simeq M_2= 50$ GeV, the heavy-light neutrino mixing
parameter is $\vert V_{\mu S_2} \vert ^{2}= 9.8 \times 10 ^{-5}$. 
Thus the heavy-light neutrino mixing is determined by $M_D$ and $M_i$ and varies inversely as
the corresponding pseudo-Dirac neutrino mass exchanged. 

\begin{figure}[h!]
\centering
\includegraphics[scale=0.58]{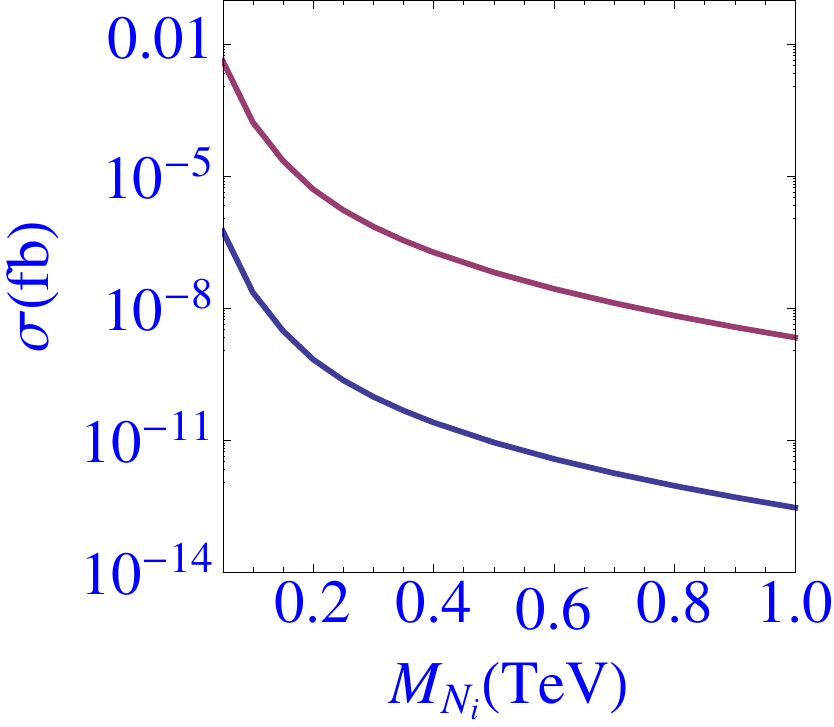}
\caption{Signal cross sections for trilepton final states 
   in the LL channel as a function of heavy pseudo Dirac mass
  $M_{N_1}$ (blue curve) and  $M_{N_2}$ (red curve) at $\sqrt s = 14$ TeV.}
\label{fig:lll12}
\end{figure}

For computation of the production cross section we have utilized the  CTEQ6M parton distribution
functions \cite{cteq} in eq.(\ref{prodcross}).   
Using our ansatz for heavy light neutrino mixing matrix $M_D/M$ in the
pseudo Dirac case, eq.(\ref{dsigdc}),
eq.(\ref{prodcross}), and eq.(\ref{inclsig}),   
our predicted results on trilepton signals in the $LL$ channel are shown for LHC energy
$\sqrt{s}=14$ TeV  in Fig.\ref{fig:lll12}   where 
$l_{(1)}l_{(2)}=e^{\pm}e^{\mp}$ and $l_{3}=e^{\pm}$ or $\mu^{\pm}$
in the lower blue curve and the mediating heavy fermion is the pseudo Dirac $N_1$ . The
corresponding trilepton signal as a function of the pseudo Dirac mass $M_{N_2}$
 is shown as upper red curve in the same figure
for which $l_{(1)}l_{(2)}=\mu^{\pm}\mu^{\mp}$ but $l_{3}=e^{\pm}$ or $\mu^{\pm}$.

\begin{figure}[h!]
\centering
\subfigure[]{%
\includegraphics[scale=0.6]{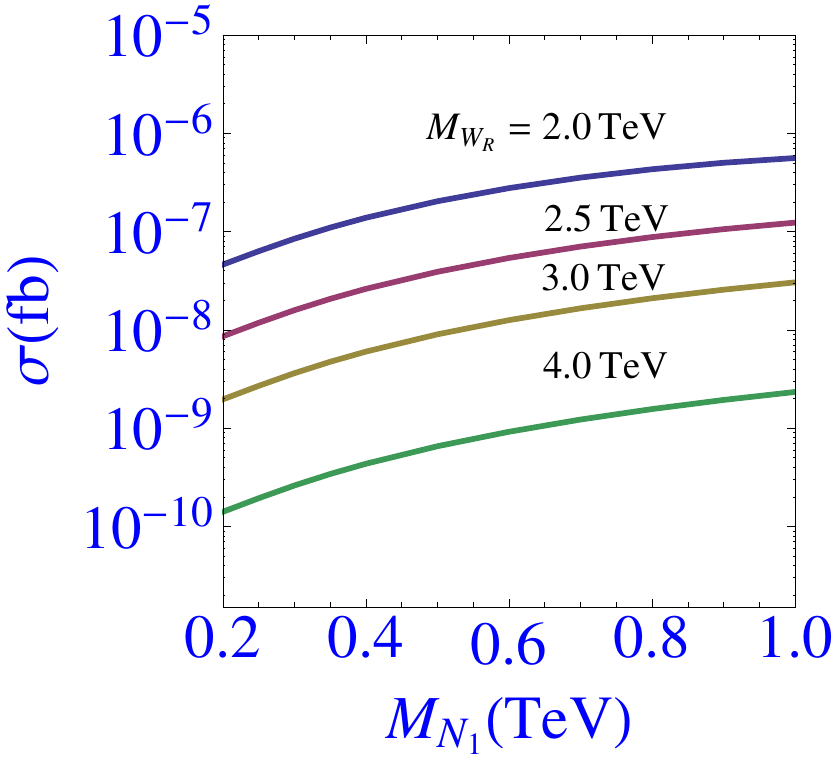}
\label{fig:rrr1}}
\quad
\subfigure[]{%
\includegraphics[scale=0.6]{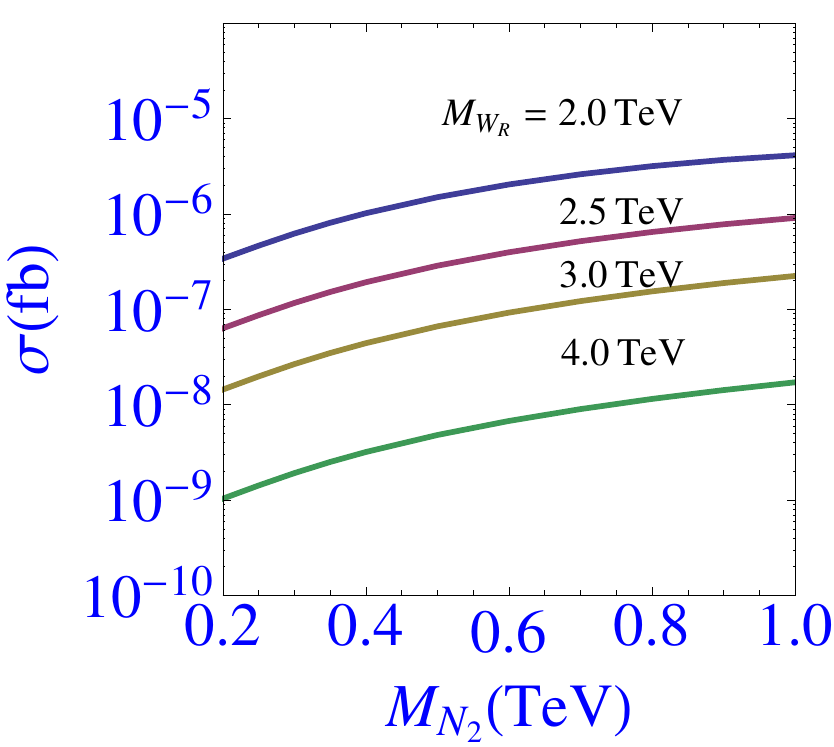}
\label{fig:rrr2}}
\caption{Same as Fig. \ref{fig:lll12} but in the $RR$ channel} 
\label{fig:figure}
\end{figure}

At $\sqrt s = 14$ TeV, the predicted trilepton signal cross sections in the $ W_R-W_R$ channel
are shown in Fig.\ref{fig:rrr1} when $l_{(1)}l_{(2)}=e^{\pm}e^{\mp}$ and
$l_{3}=e^{\pm}$ or $\mu^{\pm}$. In Fig. \ref{fig:rrr2} the predicted
signal cross sections are for 
$l_{(1)}l_{(2)}=\mu^{\pm}\mu^{\mp}$, and $l_{3}=e^{\pm}$ or
$\mu^{\pm}$ also in the same channel. In the LL channel, at 30 $fb^{-1}$ luminosity, the number of signal events for trilepton final states for  
heavy neutrino mass $M_{N}$=100 GeV is negligible. But at 3000 $fb^{-1}$ luminosity and $M_{N}$=50 GeV, the number of signal events 
becomes 12.51 indicating the presence of heavy pseudo-Dirac neutrinos. Hence, in the future run of the LHC with increased luminosity this signal may be observed and this Model-I may be verified 
or falsified. The three body decay mode of RH neutrino 
$N \rightarrow \ell W_{R}^{\star}\rightarrow \ell \ell \nu$  in the $RR$
channel is suppressed by both mixing and the heavy
 $W_{R}$ mass. 
Thus we find that the signal cross section in $LL$ channel is
dominant over that in the $RR$ channel for trilepton production at the
LHC detectors where, for a given $M_{N_i}$, the cross sections decrease
rapidly with pseudo-Dirac RH neutrino  mass. In conclusion we find that if RH neutrinos
are heavy pseudo-Dirac ($M_N > 200$ GeV), it is unlikely that LHC experiments in near
future can detect them through tri-lepton production events.

\subsection{\bf Dilepton signals at LHC detectors}

\begin{figure}[h!]
\centering
\subfigure[]{%
\includegraphics[scale=0.6]{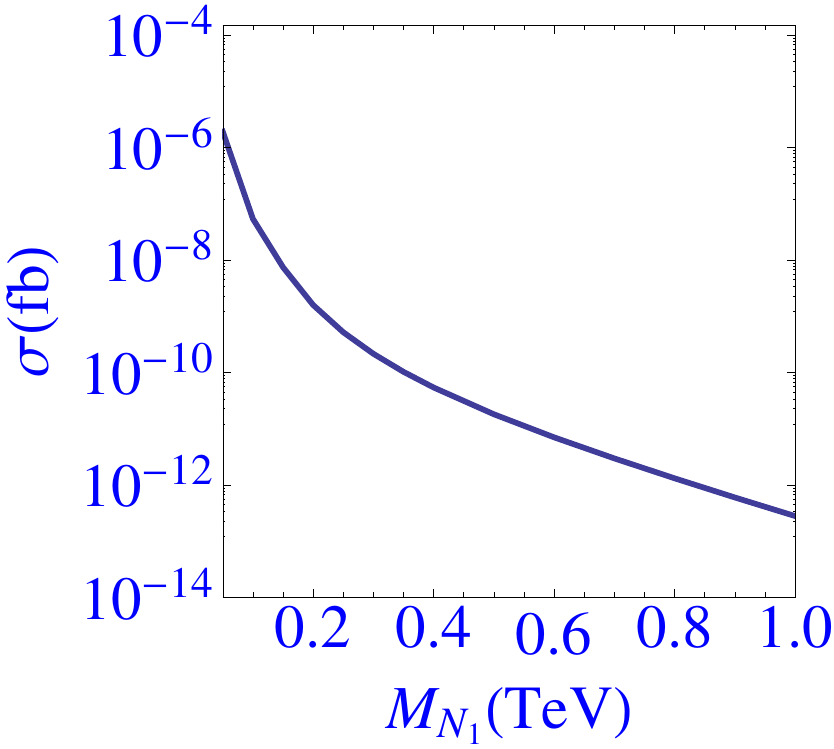}
\label{fig:ll1}}
\quad
\subfigure[]{%
\includegraphics[scale=0.6]{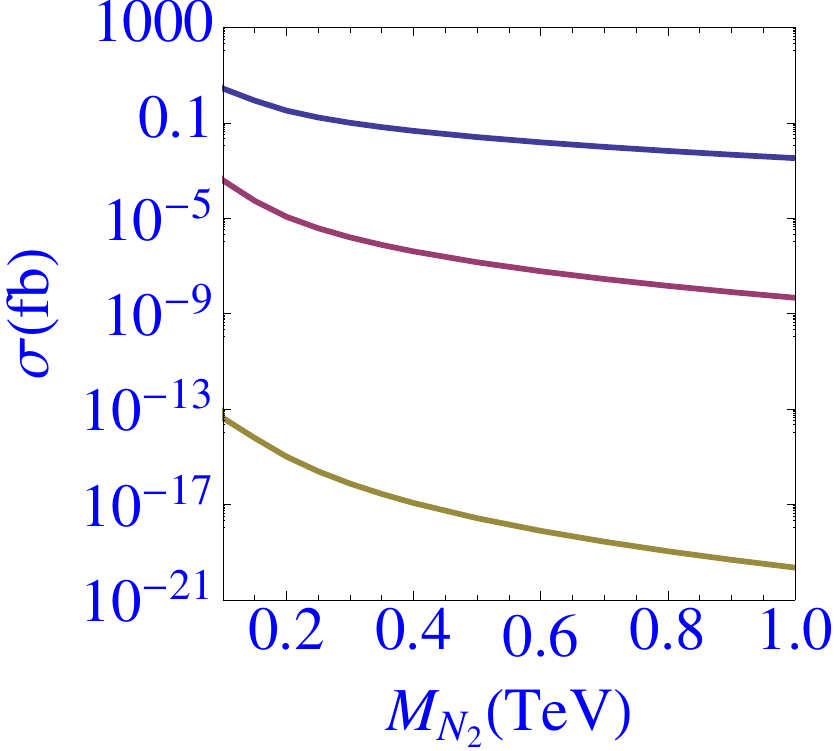}
\label{fig:ll2}}
\caption{Signal cross sections for dilepton final states in the $LL$
  channel at $\sqrt s=14$ TeV. The middle curve in the right panel represents 
the predicted signal cross section of our Model-II  while the curves above and below the middle one represent signal cross sections of 
two benchmark scenarios discussed in the text \cite{agu-saav,cdm}.}
\label{fig:dill}
\end{figure}

\begin{figure}[h!]
\centering
\subfigure[]{%
\includegraphics[scale=0.57]{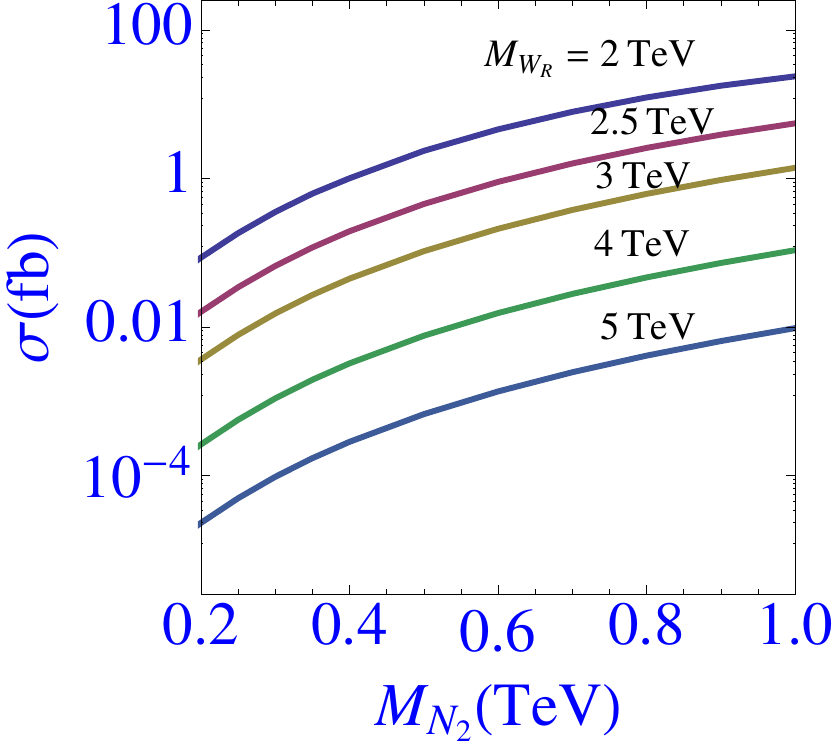}
\label{fig:dirr}}
\quad
\subfigure[]{%
\includegraphics[scale=0.55]{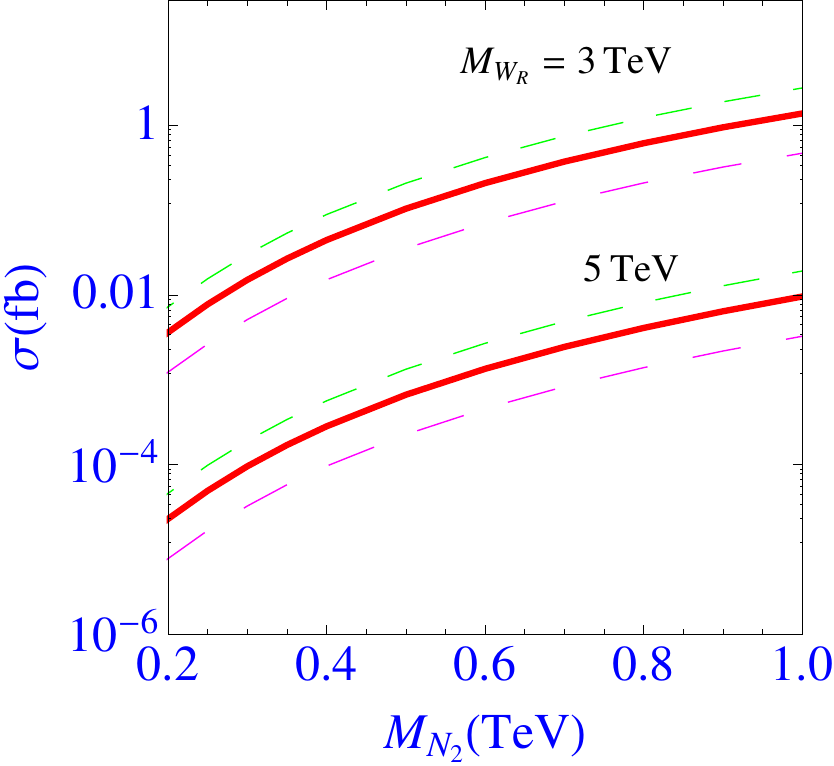}
\label{fig:diwrcom}}
\caption{(a) Predictions for dimuon signal cross section at
  $\sqrt s = 14$ TeV in the RR channel as a function of
  $(M_{W_R},M_{N_2})$ in
  Model-II with ${(g_{2L}/g_{2R})}^2= 1.24$.
(b) Comparison of different model predictions at $\sqrt s =
14$ TeV for $M_{W_R}= 3.0$ TeV (upper curves) and  $M_{W_R}= 5.0$ TeV (lower curves): (i) manifest LRS model
(green small dashed), (ii) this analysis of Model-II (solid red), and
(iii) Model of ref.\cite{app} (magenta long-dashed) for ${(g_{2L}/g_{2R})}^2= 2.4$ .} 
\label{fig:diwr}
\end{figure}

The RH Majorana neutrinos being in the fundamental representation of
LR gauge theory have direct coupling with the $W_R$ bosons which can
be produced at LHC energies manifesting in like-sign dilepton signals.
In fact, the recent CMS Collaboration has found
a lower bound $M_{W_R}\ge 3 $TeV in the manifest LRS model from their like-sign
dilepton production cross section in the $RR$ channel if the
associated RH neutrinos
are Majorana fermions \cite{rnmgs:1981}. In this experiment the 
$W_R$ boson signal is detected indirectly via like-sign dilepton
production simultaneously with two jets. 
The dilepton production process is 
significant because of the following reasons: (i) the absence of missing energy  helps in fighting the background, (ii) it is easier to reconstruct both the masses of $W_R$ and $N_R$
by measuring the energies and momenta of the final states, and (iii)
the production process can be amplified by the $W_R$ resonance.\\
In addition to three light active neutrinos, our Model-II has two
types of heavy Majorana neutrinos:\\ 
(A) Heavy RH
 neutrinos in the mass range ${\cal O}(100)$ GeV to few TeV capable of 
 mediating like-sign dilepton production inside the CMS and ATLAS
 detectors which we discuss in this section.\\
(B) Three sterile neutrinos with allowed lighter mass eigen values of ${\cal
   O}(10)$ GeV for the first or the  second generations.\\ 
 We have estimated dilepton production cross sections in Model-II in
 the $LL$, $RR$, and $RL$ channels mediated by heavy RH neutrinos at
 LHC energy of $\sqrt s= 14$ TeV. 
 The signal cross-section for the production of the RH neutrino or sterile neutrino
 including the real or virtual $W_L$,or $W_R$ exchanged at the second stage is given by 
\begin{equation}
 \sigma(pp\rightarrow Nl^{\pm}\rightarrow l^{\pm}l^{\pm}jj)=\sigma_{prod}(pp\rightarrow W_{L,R}\rightarrow Nl^{\pm})\times Br(N\rightarrow l^{\pm}jj)
\end{equation}
where the branching ratio 
\begin{equation}
 Br(N\rightarrow l^{\pm}jj)=\frac{\Gamma(N\rightarrow l^{\pm}W)}{\Gamma_{N}^{tot}}\times Br(W\rightarrow jj)
\end{equation}
here $Br(W\rightarrow jj)$=0.676\cite{pdg}.
For heavy Majorana neutrino exchange, our results are shown for
$\sqrt{s}=14$ TeV with CTEQ6M parton distribution functions in
Fig. \ref{fig:dill} and Fig. \ref{fig:dirr}, respectively, in the 
$LL$ and $RR$ channels.

In Fig.6(b), our predictions in the $RR$ channel are given by the
middle solid curve where the upper short-dashed (lower long-dashed)
curve represents our estimations in the manifest LRS model (model of
ref.\cite{app}) for two different values of $W_R$ mass, $3$ TeV , and
$5$ TeV, These predictions are subject to imposition of nearly $48\%$ cut
deduced from the conditions of LHC run-I at $\sqrt s= 8$ TeV.        
 Thus, using the predicted results of the type shown in
 Fig. 6(b), the validity of three different models can be
 tested by the LHC measurements  at $\sqrt s=14$ TeV. 
 
\begin{table}[h!]
\caption{Predictions of number of signal events for $\mu\mu jj$ as a
  function of heavy RH neutrino mass($M_{N_2}$) and luminosities($\mathcal{L}$)
  in the RR channel 
  at $\sqrt{s}$=14 TeV for $M_{W_R}$=2.5 TeV } 
\centering
\begin{tabular}{|l||l|l||l||l|l||l|}
\hline
 $M_{N_2}$ (GeV) &\multicolumn{3}{c|}{ Events  before cuts} 
 &\multicolumn{3}{c|}{ Events after cuts } \\
\cline{2-7}
 & $30 fb^{-1}$ & $300 fb^{-1}$ & $3000 fb^{-1}$ & $30 fb^{-1}$ & $300 fb^{-1}$ & $3000 fb^{-1}$\\

\hline\hline
 $200$ & $ 0.4788 $ & $4.788 $ & $ 47.88 $ & $0.2393 $ & $ 2.393$ & $23.93$\\
 $400$ & $ 5.8212 $ & $58.212 $ & $ 582.12 $& $2.9099 $ & $ 29.09$ & $290.9$\\
 $600$ & $ 26.872 $ & $268.72 $ & $ 2687.2 $& $13.452 $ & $ 134.52$ & $1345.2$\\
 $800$ & $ 77.076 $ & $770.76$ & $ 7707.6 $& $39.129 $ & $ 391.29$ & $3912.9$\\
 $1000$& $ 165.09 $ & $1650.9 $ & $ 16509 $& $89.643 $ & $ 896.43$ & $8964.3$\\
\hline
\end{tabular}
\label{tab:trr}
\end{table}

 It is observed from Fig. \ref{fig:dill} and Fig. \ref{fig:diwr} that at 30 $fb^{-1}$ luminosity, the number of signal events for 
 heavy neutrino mass $M_{N_{(1,2)}}$=100 GeV are negligible in the
 $LL$ channel, but in the  $RR$ channel these are appreciable. The
 signal events in the $RR$ channel as a function of $M_{N_2}$ and
 various luminosities are presented in Table \ref{tab:trr}. They are
 found to be more dominant compared to the LL channel where the signal
 cross sections are reduced because of damping due to heavy-light
 mixings. Such damping factors are  
 absent in the RR channel. However, the number of events in the RR channel reduces considerably
when signal cut conditions  are imposed. Even though we do not know the signal  
cut conditions at $\sqrt s = 14$ TeV, we adopt the same criteria following the latest CMS data \cite{cms2} at $\sqrt s=8$ TeV: $M_{lljj}>$ 600 GeV,
$M_{ll}>$ 200 GeV, $p_T^j>$40 GeV, $p_T^l>$40 GeV,
$p_T^{l,leading}>$60 GeV, $\left|\eta(j) \right|<$3.0 and
$\left|\eta(l) \right|<$2.5. This reduces  the number of 
signal events by nearly $48\%$. For
 example, when $M_{N_2}=800$ GeV, the number of dimuon events
are 77 (39)
 excluding (including) the effect of cuts for luminosity  $\mathcal{L}=30$ 
fb $^{-1}$.

We note that since the predicted values of  heavy-light
mixings in our model is several orders less than the upper bench mark
point (${\vert V_{lN} \vert}^2=3\times 10^{-3}$) but many orders
larger than the vanilla seesaw benchmark (${\vert V_{lN} \vert}^2=
{\sqrt (\Delta m^2_{\rm atm})}/M_N$), 
 the predicted cross sections in the LL channel
falls in between the two benchmark scenarios corresponding to the
two limits  as shown in the right-panel of Fig.\ref{fig:dill}. 

\begin{figure}[h!]
\centering
\includegraphics[scale=0.65]{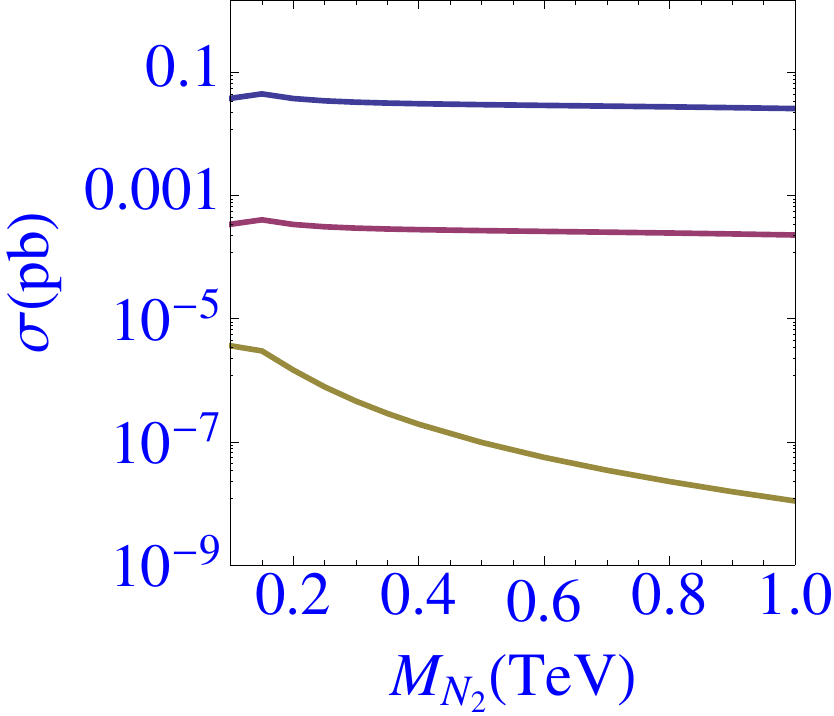}
\caption{Same as Fig.5(b) but for RL channel.}
\label{fig:dirl}
\end{figure}

Our Model-II predictions of the signal cross section in the $RL$
channel for $M_{W_R}=3$ TeV is shown by the middle curve in
Fig. \ref{fig:dirl}  which
falls below the upper curve corresponding to upper benchmark and
several orders above the vanilla seesaw benchmark. The predicted  
number of events for wider range of $M_{N_2}=100-1000$ GeV or even for larger values, excluding (including) cuts, are nearly
9(5), 28(16), 84(49) for values of proton beam luminosity $30 {\rm fb}^{-1}$, $100
{\rm fb}^{-1}$,  and  $300{\rm fb}^{-1}$, respectively. The near constancy
of observable di-muon event rates with increasing values of $M_{N_2}$
makes this channel attractive for the detection of the RH heavy
neutrino and distinguishing this channel experimentally from RR
channel which shows larger number of events with increasing behaviour.

\subsection{\bf $W_R$ boson mass from dilepton production data} 
At $\sqrt s=8$ TeV of LHC energy, the CMS collaboration \cite{cms2}
have recently observed an excess of events  in the
di-electron channel with $eejjX$ final state having a local significance
of $2.8\sigma$ at $M_{eejj}\simeq 2.1$ TeV. Here we show  how our
Model-II explains this excess.

\begin{figure}[h!]
\centering
\subfigure[]{%
\includegraphics[scale=0.33]{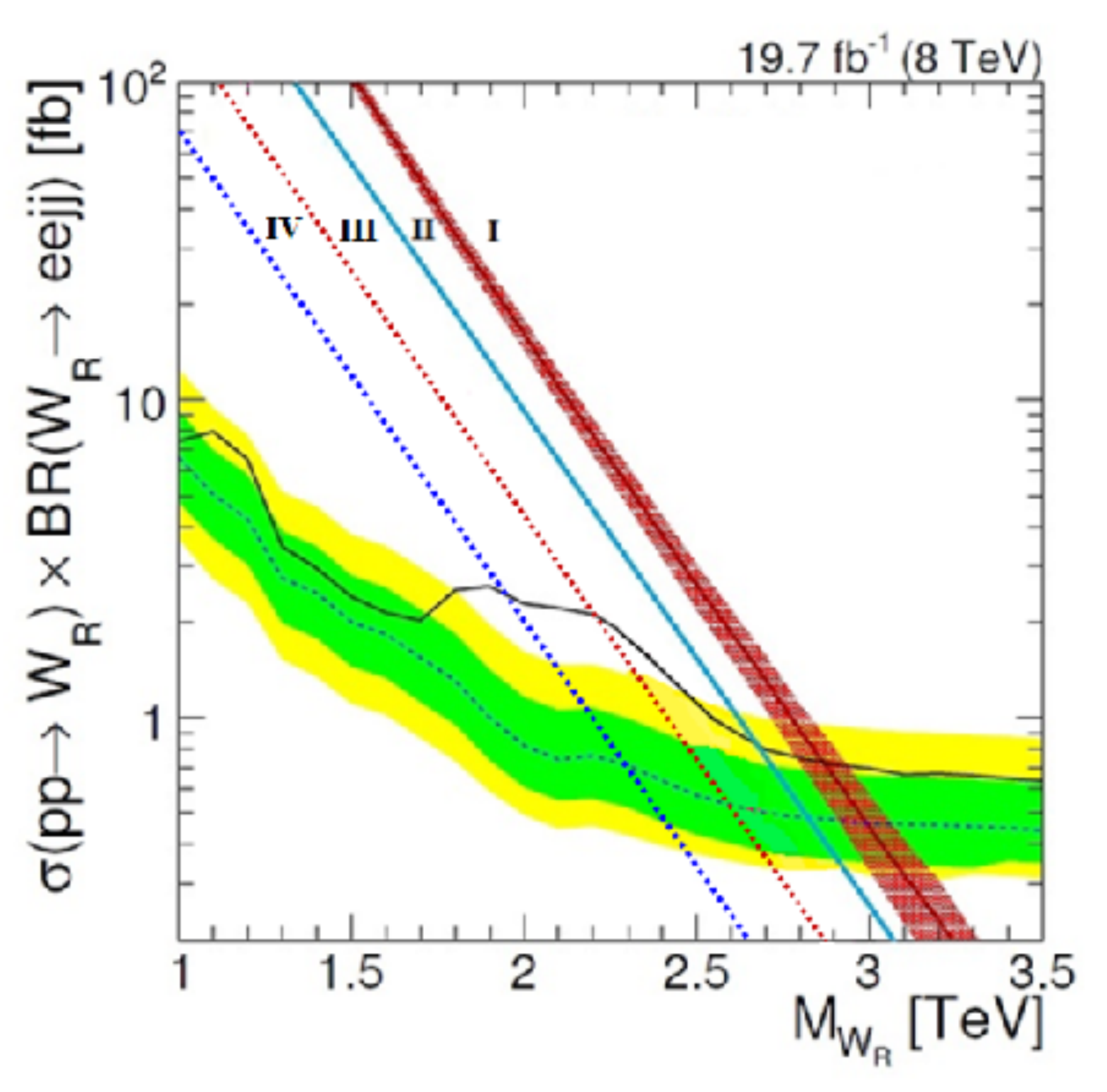}
\label{fig:CMSwree}}
\quad
\subfigure[]{%
\includegraphics[scale=0.24]{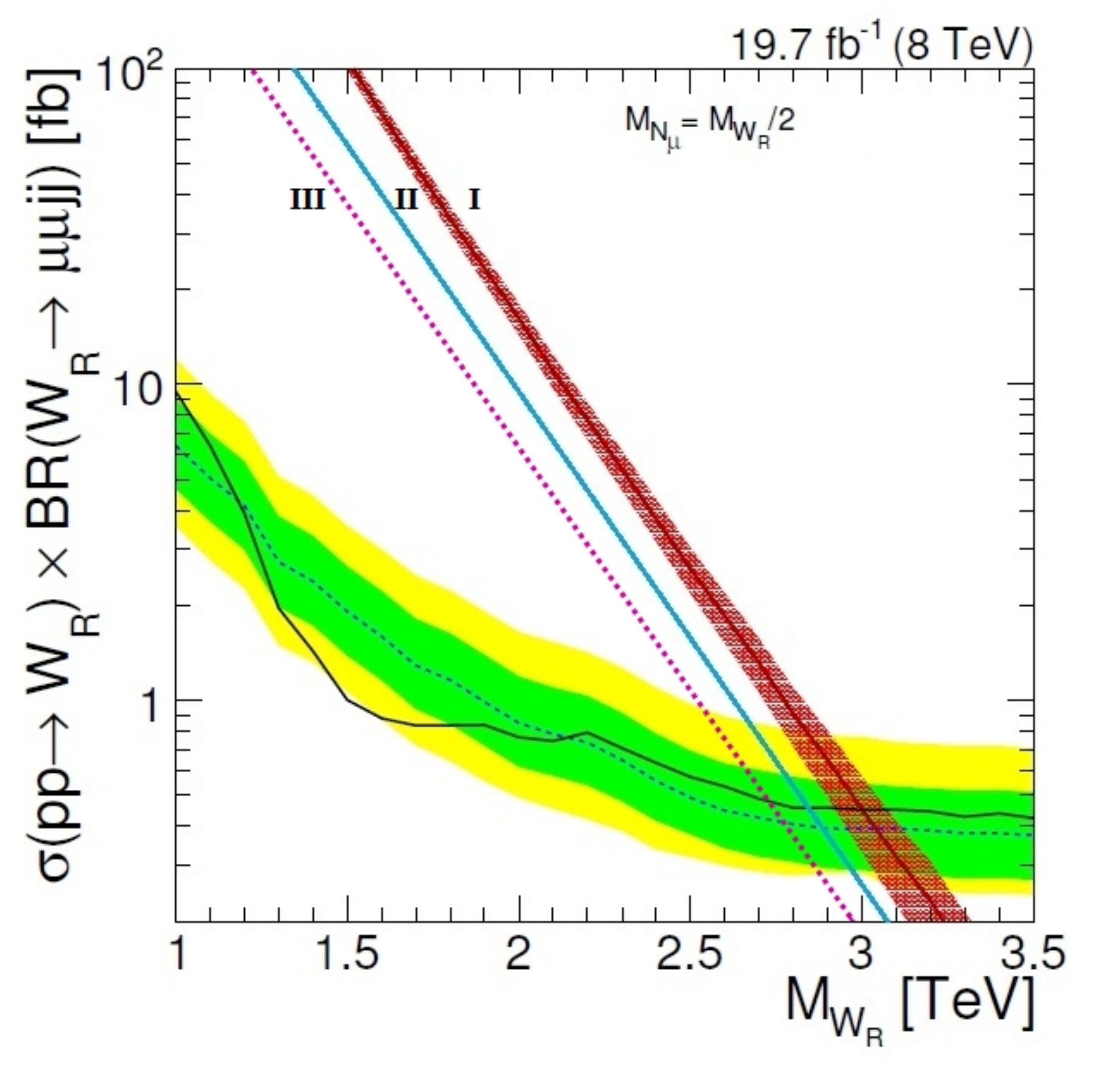}
\label{fig:CMSwrmumu}}
\caption{Predictions of like-sign di-electon (left-panel) and di-muon
  (right-panel) signal cross section shown by lines II and III  for $W_R$ production at $\sqrt s = 8$ TeV and their
  comparison with the LHC data for which the green (red) band is the $1
  \sigma$($2 \sigma$) limit. The zig-zag dotted (solid) curve
  represents expected (observed) results of measurements. The line
  III (IV) in the left-panel represents $g_{2R}=0.57$ and
  $V_{e1}^2=0.5(0.3)$. The line I in both the left and the right panels 
  with spreaded uncertainty represents the prediction of manifest LRS
  model $g_{2L}=g_{2R}$ \cite{rnmgs:1981}.}  
\label{fig:wrlhc}
\end{figure}

Using  $\sqrt s = 8$ TeV,
Model-II predictions of the dielectron and dimuon signal cross
sections are shown in the left-panel and the right panel, respectively, in
Fig. \ref{fig:wrlhc} for $W_R$ production in the $RR$ channel in comparison with the
CMS data \cite{cms2}. The line I with uncertainty band is the
prediction of the manifest LR model \cite{rnm,rnmgs:1981} for which $\sigma=1$.
The line
II is the Model-II prediction for $\sigma = (g_{2L}/g_{2R})^2= 1.24$
and $ V_{e1}^2= V_{\mu 1}^2= 1$ in both the left and the right panels
as applicable in the flavor diagonal basis of RH neutrinos. More
interesting predictions emerge in the Model-II when the RH neutrinos
are flavor non-diagonal. 
The line III represents the Model-II prediction for
the same value of $\sigma=1.24$ but for $V_{e1}^2=0.5$ (left-panel) and $V_{\mu2}^2=0.7$
(right-panel). The line IV in the left panel corresponds to
$V_{e1}^2=0.3$. In Model-II, the RH neutrino mass is heavy and does not
appear in the inverse seesaw formula that fits the neutrino
oscillation data. As such we note that our model has a wider range of
parameter space to explain  
 the observed
$eejj$ excess at $M_{W_R}\sim 2 $ TeV. Our model fits the observed
 absence of any excess of events in the $\mu\mu jj$ channel (right
 panel) for wider range of allowed values of $V_{\mu 1}^2$. A possible
 reason for the appearance of broadening of the peak around
 $M_{W_R}\sim 2$ TeV in the $eejj$ channel which has been provided in 
\cite{psb-rnm:2015} through inverse
 seesaw mechanism, seems to be applicable in the present approach also.   
Since actual experimental evidence
of $W_R$ requires a peak in the dilepton production
data with at least $5\sigma$ local significance, the observed excess
in the  $pp\to
eejjX$ channel is expected to increase in future experiments and  our
Model-II might be already indicating a smoking gun signal for the
presence of $W_R$ boson mass in the region
$M_{W_R} \sim 2.0$ TeV.

In conclusion we find that the observed excess of dilepton signal
events in the $eejj$ channel testify to the prediction of our Model-II
with $M_{W_R}=1.9-2.2$ TeV. Since the statistical significance of the
observed excess is at local significance of $2.8\sigma$, we suggest more accurate experimental
observation in this region with higher luminosity to examine if
there is such clear signal with $(5-6)\sigma$ local significance.

\section{\bf $W_R$ mass from dijet resonance and diboson signals}
In addition to the experimentally observed  excess in $pp\to eejjX$
 discussed above, the dijet
 resonance search in the $1.8$ TeV bin at  CMS\cite{CMSdij} and ATLAS
 \cite{ATLASdij} have observed excess of events at the levels of  $2\sigma$ and 
 $1\sigma$, respectively. Diboson production search has revealed
 a  $3.4\sigma$ excess for $M_{W_R}\simeq 2$ TeV at ATLAS  \cite{ATLASdb}
and a $1.4\sigma$ for $M_{W_R}\simeq 1.9$ TeV at CMS \cite{CMSdb}.  
 Further in the $1.8-1.9$ TeV bin, an excess of $2.2\sigma$  
 for $W_R\to WH$ with boosted SM Higgs boson $H$ decaying into $b{\bar b}$
 and $W \to l\nu$ has been observed \cite{CMSwh}.
All these LHC signals can be interpreted due to the production and
decay of the $W_R$ boson.

 In  LR models, the heavy $W_R$ boson which couples directly to RH
 quark-antiquark pair can be produced by the annihilation of such pair 
originating from
 the colliding proton beams. Once produced, the $W_R$ boson can mediate the dijet resonance in the
 way of producing energetic  RH
 quark-antiquark pairs through its direct coupling
 $g_{2R}{\bar q}^{\prime}_R\gamma_{\mu}q_RW_R^{\mu}$ manifesting in two
 jets.
This simple mechanism shown in the
 Feynman diagram of
 Fig.9 also provides a promising channel for the
 more direct experimental signature of $W_R$ boson at LHC compared to
 the dilepton production channel. 
\begin{figure}[h!]
\begin{center}  
\includegraphics[scale=0.5]{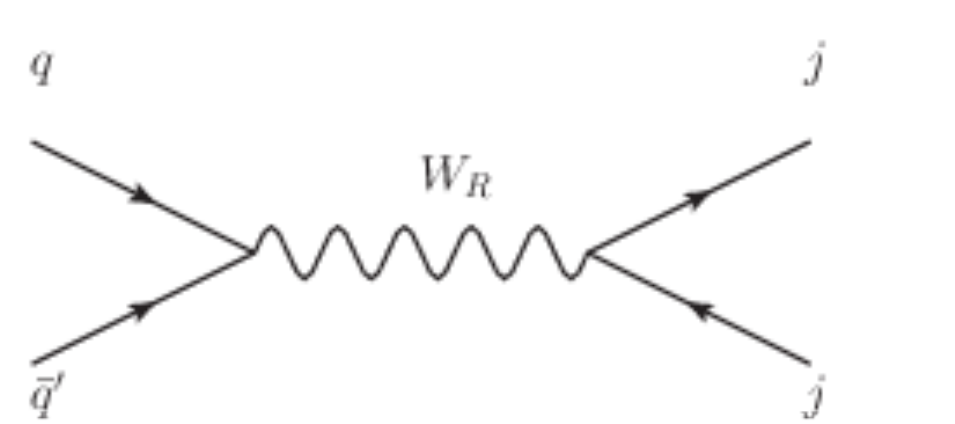}
\caption{Feynman diagram for dijet final state. }
\end{center}  
\label{fig:dijFeyn}
\end{figure}

With $g_{2R}=0.57$ in our Model-II and LHC energy $\sqrt
  s=8$ TeV
 , we predict the dijet
  production cross section $\sigma_{jj}= 288(150)$ fb for $M_{W_R}=
  1.9$ TeV 
 excluding (including) the geometric acceptance factor
  $A=0.52$ in our model . The $pp\to W_R \to WZ$ cross section is related to the
 dijet cross section \cite{bogdan:2015}
\begin{equation}
\sigma_{WZ}(W_R)=\frac{\cos^4\theta_W \eta^2_Z}{24} \sigma_{jj}(W_R),
\label{wzjjrel}
\end{equation}
leading to $\sigma_{WZ}(W_R)=3.75 {\rm fb}\times \eta^2_Z$. The ATLAS
diboson search gives  $\sigma_{WZ}(W_R) \simeq (3-10)$ fb\cite{ATLASdb}. Using
this measured cross section in the LHS  and  our
predicted value in the RHS of eq.(\ref{wzjjrel}) gives the range of
values of the parameter $0.89 <\eta_Z < 1.63$ for
$\cos^4\theta_W \simeq 0.6$. Thus, our model with $\eta_Z \sim 1$ is
consistent with the ATLAS result for $pp \to W_R\to WZ$.

In the other diboson search channel corresponding to $pp \to W_R\to WH$,\\  
$\sigma_{WH}(W_R)\approx {\sigma_{WZ}(W_R)}/{\cos^4\theta_W }$
which gives $\sigma_{WH}(W_R) \sim 6$ fb for $M_{W_R} =1.9$ TeV in our
case consistent with the CMS experimental upper bound
${[\sigma_{WH}(W_R)]}_{CMS} <18$ fb at $M_{W_R}=1.8$ TeV.   
 With  $g_{2R}=0.56$ in our Model-I, the predictions for dijet and
 diboson decay channels for $W_R$ are similar. Needless to mention
 that the dijet and diboson production results for $W_R$ are independent of the
 nature of RH neutrino (pseudo-Dirac or Majorana). 

\par\noindent{\bf Summary :} In summary including Planck-scale effects
induced by a non-renormalizable ${\rm dim.}5$ operator in $SO(10)$
through the ${210}_H$ representation and incorporating the
fine-structure constant matching condition and GUT threshold effects,
we have shown the realization
of LHC scale LR gauge theory in the minimal chain with minimal light
Higgs spectrum in concordance  with neutrino oscillation data through
experimentally verifiable gauged inverse seesaw mechanism that
predicts TeV scale heavy neutrinos  either as pseudo-Dirac (Model-I)
fermions manifesting through tri-lepton production
 or as Majorana (Model-II)  fermions manifesting as like-sign dilepton
 production signals at the LHC. The existence of
 LR gauge theory covers the predicted range of the mass scale $M_R \sim 10^3 - 10^5$ GeV with experimentally measurable
proton lifetimes.
The heavy-light neutrino mixings are predicted via charged fermion
mass fits and the charged LFV constraints consistent with  branching
ratios only few to four orders smaller than the current experimental limits.
The Model-II  permits at least one light sterile neutrino
that mediates dominant $0\nu\beta\beta$ decay rate in the $W_L-W_L$ channel irrespective of the light neutrino mass hierarchies  and independent of other possible contributions  through $W_L-W_R$ mixings.
Both the models are found to be consistent with dijet and 
$W_R\to WZ$, and $W_R \to WH$ production data for masses of  
$M_{W_R}\simeq 2$ TeV. In Model-II
the resonant production of 
$W_R$ boson and its subsequent decay  in the $RR$ channel through the
heavy RH neutrino are found to explain the recently observed
excess of events in  $pp \to  eejj X$ at the CMS detector 
predicting its mass range $M_{W_R}= 1.9-2.2$ TeV which is also
consistent with the value obtained from dijet resonance and diboson production
data. The model has  also the potential of
explaining the baryon asymmetry of the universe via resonant leptogenesis mediated by the
${\cal O} (500) $ GeV quasi-degenerate masses of the second and the
third generation sterile neutrinos noted recently\cite{bpnmkp} which would be
investigated elsewhere \cite{bismkp}. 
 Only for gauge
coupling unification in the pseudo Dirac 
case, the Model-I has just one bidoublet and one
RH doublet carrying $B-L=-1$. In
Model-II, when all neutral fermions are Majorana particles,
 there is just one more RH triplet Higgs scalar carrying
$B-L=-2$  at the LHC scale. The singlet fermions can be embedded into non-standard
fermion representation ${45}_F\subset SO(10)$. These Higgs masses are
accessible to LHC and future colliders where experimental tests 
 can discriminate this model
from others.
In conclusion we note that the Model-II has  high degree of falsifiability
 from its rich structure of verifiable predictions. In order to test both the
 Model-I and Model-II with much better accuracy, LHC data at higher
 luminosity at $\sqrt s= 8$ TeV,  and $\sqrt s= 13-14$  TeV are
 necessary.  Our estimation in the $RR$ channel at LHC run-II
 for $\sqrt s=14$ TeV predicts dijet production cross sections 
   nearly $6$ times larger than its current value. 

\par\noindent{\bf ACKNOWLEDGMENT:-}
M. K. P. thanks the Science and Engineering Research Board, Department of Science and Technology, Govt. of India
for the research project SB/S2/HEP-011/2013. B. S. thanks SOA
University for a research fellowship. The authors thank Ram Lal Awasthi and
Samiran Bose for computational help.

\section*{References}

\bibliography{mybibfile}

\end{document}